# Tidal Deformability of Neutron Stars in Scalar-tensor Theories of Gravity

Stephanie M. Brown[1,2]
[1] Max-Planck-Institut für Gravitationsphysik (Albert-Einstein-Institut), Callinstraße 38, D-30167 Hannover, Germany; stephanie.brown@fysik.su.se
[2] Leibniz Universität Hannover, D-30167 Hannover, Germany
*Received 2022 November 30; revised 2023 September 14; accepted 2023 September 15; published 2023 November 17*

## Abstract

Gravitational waves from compact binary coalescences are valuable for testing theories of gravity in the strong field regime. By measuring neutron star tidal deformability using gravitational waves from binary neutron stars, stringent constraints were placed on the equation of state of matter at extreme densities. Tidal Love numbers in alternative theories of gravity may differ significantly from their general relativistic counterparts. Understanding exactly how the tidal Love numbers change will enable scientists to untangle physics beyond general relativity from the uncertainty in the equation of state measurement. In this work, we explicitly calculate the fully relativistic $l \geqslant 2$ tidal Love numbers for neutron stars in scalar-tensor theories of gravitation. We use several realistic equations of state to explore how the mass, radius, and tidal deformability relations differ from those of general relativity. We find that tidal Love numbers and tidal deformabilities can differ significantly from those in general relativity in certain regimes. The electric tidal deformability can differ by ∼200%, and the magnetic tidal deformability differs by ∼300%. These deviations occur at large compactnesses ($C = M/r \gtrsim 0.2$) and vary slightly depending on the equation of state. This difference suggests that using the tidal Love numbers from general relativity could lead to significant errors in tests of general relativity using the gravitational waves from binary neutron star and neutron star black hole mergers.

## 1. Introduction and Motivation

Compact objects such as neutron stars and black holes are essential for testing general relativity (GR) in the strong field regime. Gravitational waves (GWs) emitted by compact objects by LIGO-Virgo have improved our understanding of gravity in the strong field regime. The LIGO/Virgo collaboration has detected almost one-hundred compact binary coalescences to date: two binary neutron star mergers, two neutron star black hole mergers, and more than eighty binary black hole mergers (Abbott et al. 2019a, 2021a, 2021c). An independent analysis of the available data found even more events (Nitz et al. 2020, 2021a, 2021b). Previous analyses of these events have already placed limits on possible deviations from GR (Abbott et al. 2019b, 2021b; Wang et al. 2021, 2022; Mehta et al. 2022). Recently, waveforms for various alternate theories of gravity have been developed and applied to parameter estimation. These waveforms allow for stringent tests of various theories of gravity. They also enable more general tests for physics beyond GR such as placing limits on the existence of scalar and tensor propagation modes (Mirshekari & Will 2013; Nair et al. 2019; Chatziioannou et al. 2021).

Neutron stars are also unique laboratories for studying nuclear physics at ultra-high densities. Information about neutron star matter is encoded in GWs from binary neutron star and neutron star black hole mergers (Hebeler et al. 2010, 2013; Özel & Freire 2016). Neutron stars contain vital information needed to understand phases of matter encountered in quantum chromodynamics. The tidal deformability encodes information about the nuclear equation of state in GWs (Hinderer 2008; Binnington & Poisson 2009; Damour & Nagar 2009). Studies of binary neutron star merger GW170817 have improved our knowledge of the nuclear equation of state (Abbott et al. 2018; Radice & Dai 2019; Capano et al. 2020; Raaijmakers et al. 2021). Despite this, the nuclear equation of state is still unknown. Studying neutron stars in alternative theories of gravity is challenging because deviations in neutron star properties caused by non-GR effects are of the same order of magnitude as the uncertainty in the equation of state. Understanding how the mass–radius–tidal deformability relationships deviate from GR is essential to untangling these differences.

Tidal deformability connects GWs and the nuclear equation of state. Tidal deformabilities and the associated tidal Love numbers relate an applied external tidal field to the induced internal multipole moment, measuring the magnitude of deformation under a given tidal force. Love numbers were initially defined in Newtonian gravity (Love 1909; Shida 1912) and then expanded to GR by Flanagan & Hinderer (2008), Hinderer (2008). The concept was further expanded and made more concrete in several follow-up papers, including Binnington & Poisson (2009), Damour & Nagar (2009).

This work focuses on scalar-tensor theory, one of the most natural and best studied alternate theories of gravity. The theory was initially motivated partly by Mach's principle (Brans & Dicke 1961) and partly in an attempt to expand GR to five dimensions (Jordan 1955). However, it is still of interest today. Scalar degrees of freedom are critical for string theory, superstring theory, and other supergravity theories (Fujii & Maeda 2003). Therefore, scalar-tensor theories can sometimes be used as a phenomenological proxy for more complex extensions of GR. Furthermore, scalar fields have been proposed as an alternative solution to the dark energy problem (Garcí-Bellido & Quirós 1990; Boisseau et al. 2000; Clifton et al. 2012).

Scalar-tensor theories add a massless scalar field ($\varphi$) to the standard GR metric ($g_{\mu\nu}$). The metric and the scalar field are coupled into an *effective metric* $\tilde{g}_{\mu\nu} = A^2(\varphi)g_{\mu\nu}$. The earliest versions of this theory were presented more than half a century ago by Fierz (1956), Jordan (1955), and Brans & Dicke (1961). In the simplest scalar-tensor theory, known as FJBD (Fierz, Jordan, and Brans and Dicke), the scalar field is coupled to the metric by the coupling function $A(\varphi) = e^{\alpha\varphi}$. Solar system experiments have placed stringent constraints on the value of $\alpha$ (Shapiro 1990). These constraints also significantly limit the strong-field behavior. Damour and Esposito-Ferèse discovered the "spontaneous scalarization" effect, which allows large deviations from GR in the strong field regime without violating the strict solar system constraints. Damour and Esposito-Ferèse defined $A(\varphi) = e^{\beta\varphi^2/2}$ and found that scalarization occurs for $\beta \lesssim -4.5$ (Damour & Esposito-Farèse 1993). A follow-up study showed that scalarization occurs for $\beta \lesssim -4.35$ (Harada 1998).

In this work, we calculate the tidal Love numbers of neutron stars in scalar-tensor theories of gravity, focusing on the spontaneous scalarization case. Section 2 presents the equilibrium configuration for neutron stars in scalar-tensor theory. Section 3 discusses the first-order linear time-independent perturbations upon which the tidal deformabilities depend. Section 4 details the method for deriving the various tidal Love numbers. Section 5 presents the results and demonstrates how the Love numbers in scalar-tensor theories differ from those in GR. The paper concludes with Section 6, which discusses the results.

## 2. Neutron Stars in Scalar-tensor Theory

Scalar-tensor theories are straightforward alternatives to GR. They depend on both a metric tensor ($g_{\mu\nu}$) and a massless scalar field ($\varphi$) and are typically expressed in one of two conformal frames: the Einstein frame and the Jordan frame. Historically, there has been much debate over the correct choice of frame (Deruelle & Sasaki 2011), but it is now agreed that experiments measure Jordan frame quantities even though the field equations simplify in the Einstein frame (Barausse et al. 2013; Doneva et al. 2013; Palenzuela et al. 2014; Pani & Berti 2014; Crisostomi et al. 2018).

In the Jordan frame, the action is

$$S = \frac{1}{16\pi G}\int\sqrt{-\tilde{g}}\left(\phi\tilde{R} - \frac{\omega(\phi)}{\phi}\tilde{g}^{\mu\nu}\partial_\mu\phi\partial_\nu\phi + 2\lambda(\phi)\right)d^4x + S_m[\Psi_m, \tilde{g}_{\mu\nu}], \tag{1}$$

where the tilde denotes Jordan frame quantities, $\phi$ is the Jordan frame scalar field, $\tilde{g}_{\mu\nu}$ is metric, $\tilde{R}$ is the Ricci scalar, $\omega(\phi)$ is a function of the scalar field that characterizes a specific scalar-tensor theory, and $\lambda(\phi)$ is the scalar potential. $S_m$ denotes the action of the matter, which is a function of the matter fields $\Psi_m$ and the Jordan metric $\tilde{g}_{\mu\nu}$. Due to the $\phi\tilde{R}$ term, the gravitational constant $G$ becomes a function of the scalar field, i.e., $\tilde{G} = G(\phi)$. Throughout this work, we will continue to denote Jordan frame quantities with a tilde.

The Jordan frame is the physical frame, but the field equations are typically expressed in the Einstein frame, where the metric and scalar decouple. A conformal transformation relates the two frames:

$$\tilde{g}_{\mu\nu} = A^2(\varphi)g_{\mu\nu}. \tag{2}$$

Using this transformation, the action can be rewritten in a way that resembles the Einstein–Hilbert action:

$$S = \frac{1}{16\pi G_*}\int\sqrt{-g}\,(R - 2g^{\mu\nu}\partial_\mu\varphi\partial_\nu\varphi - 2\lambda(\varphi))d^4x + S_m[\Psi_m, A^2(\varphi)g_{\mu\nu}] \tag{3}$$

where all quantities are related to the Einstein metric $g_{\mu\nu}$. $\varphi$ is the Einstein frame scalar field, $R$ is the scalar curvature, and $G_*$ is the bare gravitational coupling constant, which is set to 1, along with $c$, from here on. This paper will focus on the $\lambda(\varphi) = 0$ case.

The Jordan ($\phi$) and Einstein ($\varphi$) frame scalar fields are related by the following equation (Palenzuela et al. 2014):

$$\phi = e^{-\beta\varphi^2}. \tag{4}$$

Much of the work presented here is applicable for any $A(\varphi)$, but when necessary, the spontaneous scalarization coupling function (Damour & Esposito-Farèse 1993) is used:

$$A(\varphi) = e^{\beta\varphi^2/2}. \tag{5}$$

The modified field equations, derived from the Einstein frame action, have the form

$$G_{\mu\nu} = 8\pi G_* T_{\mu\nu} + T^{(\varphi)}_{\mu\nu}; \tag{6a}$$

$$\Box\varphi = -4\pi G_*\alpha(\varphi)T, \tag{6b}$$

where $\alpha(\varphi) \equiv d\ln A(\varphi)/d\varphi$. $T^{(\varphi)}_{\mu\nu}$ can be considered the stress-energy of the massless scalar field and has the form

$$T^{(\varphi)}_{\mu\nu} \equiv 2\partial_\mu\varphi\partial_\nu\varphi - g_{\mu\nu}g^{\alpha\beta}\partial_\alpha\varphi\partial_\beta\varphi. \tag{7}$$

$T_{\mu\nu}$ is the stress-energy tensor in the Einstein frame, and $T$ is the contracted stress-energy tensor $T = T^\mu_\mu = g^{\mu\nu}T_{\mu\nu}$. $T_{\mu\nu}$ is related to the Jordan frame stress-energy tensor ($\tilde{T}_{\mu\nu}$) in the following manner

$$T^{\mu\nu} \equiv \frac{2}{\sqrt{|g|}}\frac{\delta S_m}{\delta g_{\mu\nu}} = A^6(\varphi)\tilde{T}^{\mu\nu}. \tag{8}$$

Note setting $\alpha(\varphi)$ to zero retrieves the GR field equations.

We model neutron stars as static, spherically symmetric, nonrotating objects and assume that neutron star matter can be described as a perfect fluid. The stress-energy tensor for a perfect fluid is defined in the physical frame as

$$\tilde{T}_{\mu\nu} = (\tilde{\rho} + \tilde{p})\tilde{u}_\mu\tilde{u}_\nu - \tilde{p}\tilde{g}_{\mu\nu} \tag{9}$$

where $\tilde{u}_\mu$ is the four-velocity of the fluid, and $\tilde{\rho}$ and $\tilde{p}$ are the energy density and pressure in the Jordan frame. We assume that $\tilde{p}$ and $\tilde{\rho}$ are related by some barotropic equation of state so that

$$\delta\tilde{\rho} = \frac{d\tilde{\rho}}{d\tilde{p}}\delta\tilde{p}, \tag{10}$$

where $\delta\tilde{p}$ and $\delta\tilde{\rho}$ are the Eulerian fluid perturbations. As the star is static, only the $t$ component of the four-velocity is nonzero:

$$u^\mu = (e^{\nu/2}, 0, 0, 0). \tag{11}$$

Conservation of energy and momentum is defined in the physical or Jordan frame, i.e., $\tilde{\nabla}_\mu\tilde{T}^\mu_\nu = 0$. Transforming to the

Einstein frame gives

$$\nabla_\nu T^\nu_\mu = \alpha(\varphi) T \, \nabla_\mu \varphi. \tag{12}$$

The metric for a static, spherically symmetric, self-gravitating object is

$$ds^2 = g_{\alpha\beta} dx^\alpha dx^\beta = -e^\nu dt^2 + e^\lambda dr^2 + r^2(d\theta^2 + \sin^2\theta d\phi^2) \tag{13}$$

where $\nu$ and $\lambda$ are functions of $r$ and $e^{-\lambda} = 1 - 2\mu(r)/r$.

The modified Tolman–Oppenheimer–Volkoff (TOV) or structure equations, which can be derived from the field equations and the equation for conservation of energy, have the form

$$\frac{d\mu}{dr} = 4\pi G_* r^2 A^4(\varphi)\tilde{\rho} + \frac{1}{2} r(r - 2\mu)\psi^2; \tag{14a}$$

$$\frac{d\nu}{dr} = 8\pi G_* \frac{r^2 A^4(\varphi)\tilde{p}}{r - 2\mu} + r\psi^2 + \frac{2\mu}{r(r - 2\mu)}; \tag{14b}$$

$$\frac{d\varphi}{dr} = \psi; \tag{14c}$$

$$\frac{d\psi}{dr} = 4\pi G_* \frac{rA^4(\varphi)}{r - 2\mu}[\alpha(\varphi)(\tilde{\rho} - 3\tilde{p}) + r\psi(\tilde{\rho} - \tilde{p})] - \frac{2(r - \mu)}{r(r - 2\mu)}\psi; \tag{14d}$$

$$\frac{d\tilde{p}}{dr} = -(\tilde{\rho} + \tilde{p})\left[4\pi G_* \frac{r^2 A^4(\varphi)\tilde{p}}{r - 2\mu} + \frac{1}{2} r\psi^2 + \frac{\mu}{r(r - 2\mu)} + \alpha(\varphi)\psi\right] \tag{14e}$$

where $\mu$ is the mass function. $\psi = \partial_r \varphi$ is used throughout this paper for improved readability.

## 3. Stationary Perturbations

In this section, we compute the linear, time-independent scalar and spacetime perturbations following the method initially laid out by Thorne & Campolattaro (1967). The complete system of time-dependent perturbations in scalar-tensor theory was calculated in Sotani & Kokkotas (2005), and the perturbation equations in this section have been cross checked with the extant results.

We use the Regge–Wheeler gauge (Regge & Wheeler 1957), which separates the metric perturbation $h_{\mu\nu}$ into its even and odd parity components $h_{\mu\nu} = h^+_{\mu\nu} + h^-_{\mu\nu}$. Sotani & Kokkotas (2005) demonstrated that the metric in both frames can be written in the Regge–Wheeler gauge using the proper redefinition of the metric components between frames.

For this analysis, as we are interested in time-independent perturbations, all perturbations ($H_0$, $H_2$, $K$, $h_0$, and $h_1$) are functions of $r$ only. Furthermore, the $tr$ term ($H_1$) that is typically present in the Regge–Wheeler gauge vanishes.

The Einstein metric can be written in the following way:

$$h_{\mu\nu} = h^+_{\mu\nu} + h^-_{\mu\nu}; \tag{15}$$

where

$$h^+_{\mu\nu} = \sum_{\ell=2}^{\infty}\sum_{m=-\ell}^{\ell} \begin{bmatrix} e^\nu H_{0,\ell m} & 0 & 0 & 0 \\ 0 & e^\lambda H_{2,\ell m} & 0 & 0 \\ 0 & 0 & r^2 K_{\ell m} & 0 \\ 0 & 0 & 0 & r^2 K_{\ell m}\sin^2\theta \end{bmatrix} Y_{\ell m}(\theta, \phi); \tag{16}$$

and

$$h^-_{\mu\nu} = \sum_{\ell=2}^{\infty}\sum_{m=-\ell}^{\ell} \begin{bmatrix} 0 & 0 & -h_{0,\ell m}\sin^{-1}\theta\partial_\phi & h_{0,\ell m}\sin\theta\partial_\theta \\ 0 & 0 & -h_{1,\ell m}\sin^{-1}\theta\partial_\phi & h_{1,\ell m}\sin\theta\partial_\theta \\ \text{sym} & \text{sym} & 0 & 0 \\ \text{sym} & \text{sym} & 0 & 0 \end{bmatrix} Y_{\ell m}(\theta, \phi); \tag{17}$$

where $Y_{\ell m}(\theta, \phi)$ is the spherical harmonic function for $l$, $m$, and sym indicates that the metric is symmetric.

The explicit form of the conformal transformation between the Jordan and Einstein frame perturbation ($\tilde{h}_{\mu\nu} \to h_{\mu\nu}$) is needed to determine the Jordan frame tidal deformability. The conformal transformation is obtained by perturbing Equation (2) and substituting in the Regge–Wheeler metric (Sotani & Kokkotas 2005). This gives

$$h_{\mu\nu} = \frac{1}{A^2(\varphi)}\tilde{h}_{\mu\nu} - \frac{2}{A(\varphi)} g_{\mu\nu}\delta A \tag{18}$$

where $\delta A$ is the variation of the conformal factor; it is a function of the scalar field perturbation $\delta\varphi = \delta\varphi(r)$. The relationship between $\delta A$ and $\delta\varphi$ depends on the functional form of the conformal factor. In the case of spontaneous scalarization $\delta A = \beta A(\varphi)\varphi\delta\varphi$. The explicit relationships between the individual metric perturbations are>

$$\tilde{H}_0 = A^2(\varphi)H_0 - 2A(\varphi)\delta A; \tag{19a}$$

$$\tilde{H}_2 = A^2(\varphi)H_2 + 2A(\varphi)\delta A; \tag{19b}$$

$$\tilde{K} = A^2(\varphi)K + 2A(\varphi)\delta A; \tag{19c}$$

$$\tilde{h}_{0,1} = A^2(\varphi)h_{0,1}. \tag{19d}$$

We dropped the $\ell m$ subscripts from $H_0$, $H_2$, $K$, $h_0$, and $h_1$ for readability and will continue to do so throughout this work.

The complete set of perturbation equations needed to calculate the tidal deformability are laid out in Appendix B.

In GR, the full system of time-independent perturbed equations can be reduced to one differential equation for each parity: one for the even parity tensor perturbation ($H$) and one for the odd parity tensor perturbation ($h$). Scalar-tensor theories have an additional equation for the scalar field, which is of even parity. The metric and the scalar field are decoupled in the Einstein frame; therefore, the equations for $H$ and $\delta\varphi$ decouple. The Jordan frame perturbation $\tilde{H}$ depends on both $H$ and $\delta\varphi$.

## 4. Neutron Star Tidal Deformability

We derive and compute the scalar-tensor tidal Love numbers and tidal deformabilities using the method developed by Hinderer (2008) and extended in Binnington & Poisson (2009), Damour & Nagar (2009).

Tidal deformabilities (e.g., $\lambda$) relate an applied external tidal field ($\mathcal{E}_{ij\ldots k}$) to the induced multipole moment ($Q_{ij\ldots k}$). To linear order in $\mathcal{E}_{ij\ldots k}$, the tidal deformability is a proportionality constant between the two (Hinderer 2008), i.e.,

$$Q_{ij\ldots k} = -\lambda\mathcal{E}_{ij\ldots k}. \tag{20}$$

Both $\mathcal{E}_{ij\ldots k}$ and $Q_{ij\ldots k}$ can be decomposed into tensor harmonics

$$\mathcal{E}_{ij\ldots k} = \sum_{m=-\ell}^{\ell} \mathcal{E}_m Y^{\ell m}_{ij\ldots k}(\theta, \phi); \tag{21}$$

$$Q_{ij\ldots k} = \sum_{m=-\ell}^{\ell} Q_m Y^{\ell m}_{ij\ldots k}(\theta, \phi) \tag{22}$$

where $Y^{\ell m}_{ij\ldots k}(\theta, \phi)$ are the even parity tensor spherical harmonics defined by Thorne (1980). This means that the tensor relation in Equation (20) can be expressed as a scalar relation

$$Q_m = -\lambda \mathcal{E}_m. \tag{23}$$

To calculate $\lambda$, it is sufficient to calculate one nonvanishing $\mathcal{E}_m$ (Hinderer 2008).

A scalar tidal deformability $\lambda^{(\varphi)}$ is defined analogously, i.e.,

$$Q^{(\varphi)}_{ij\ldots k} = -\lambda^{(\varphi)} \mathcal{E}^{(\varphi)}_{ij\ldots k} \tag{24}$$

where $Q^{(\varphi)}_{ij\ldots k}$ and $\mathcal{E}^{(\varphi)}_{ij\ldots k}$ are the scalar tidal and multipole terms.

The external tidal field and the induced multipole moment affect spacetime in and around the neutron star. Outside the star, the large $r$ behavior of the metric can be written in terms of $\mathcal{E}_{ij}$ and $Q_{ij}$ (Thorne 1998; Hinderer 2008). For example, the metric expansion for a spherically symmetric star of mass $\mu$ in a quadrupolar tidal field $\mathcal{E}_{ij}$ for large $r$ is

$$g_{tt} = g^0_{tt} + h_{tt} = -\left(1 - \frac{2\mu}{r}\right) + \frac{3Q_{ij}}{r^3}\left(n^i n^j - \frac{1}{3}\delta^{ij}\right) + \mathcal{O}(r^{-4}) - \mathcal{E}_{ij} n^i n^j r^2 + \mathcal{O}(r^3) \tag{25}$$

where $n^i$ is the unit radial vector.

### 4.1. Electric Type Love Numbers

In GR, the electric or even parity Love numbers are calculated from the $g_{tt} = g^{(0)}_{tt} + h_{tt}$ component of the metric and are based on a single, second-order linear differential equation, for $H = H_0 = H_2$. However, in scalar-tensor theory, there are two types of even parity perturbations: scalar and tensor. The even parity metric tidal Love numbers $k_\ell$ define how the body responds to a change in the *metric*. The scalar tidal Love numbers $\kappa_\ell$ define how the body responds to a change in the applied *scalar field*. As the scalar field and the metric are not coupled in the Einstein frame, a change in the matter field does not induce a scalar perturbation, and vice versa.[3] The perturbation equations for the tidal Love number calculation must be derived carefully to the first order in either the scalar perturbation or the metric perturbation but not both. This approach differs from previous approaches in Yazadjiev et al. (2018), Pani & Berti (2014).

There are two master equations: one second-order linear differential equation for the tensor perturbation $H = H_0 = H_2$, which comes from the perturbation of the field Equation 6(a), and one for the scalar perturbation $\delta\varphi$, which comes from the scalar wave equation (Equation 6b(b)).

While the differential equation for $\varphi$ can be derived directly from the scalar wave equation (see Equation (B26)), the differential equation for $H$ is derived from the a system of Equations (B20) to (B25) and is obtained by the following steps (which have been widely used in GR Hinderer 2008; Binnington & Poisson 2009; Damour & Nagar 2009):

1. Equation (B20) $\longrightarrow H_0 = H_2 \equiv H$;
2. Equation (B21) $\longrightarrow K' = H' + \nu' H$;
3. Equation (B22) $\longrightarrow K'' = H'' + \nu' H' + \nu'' H$;
4. Equation (B24) $\longrightarrow \delta p = \frac{1}{2}(p + \rho)H$;
5. Equation (B25) $\longrightarrow H'' + c_1 H' + c_2 H = 0$.

This gives

$$H'' + \left(\frac{2}{r} + \frac{1}{2}(\nu' - \lambda')\right)H' - \left(e^{\lambda}\left(\frac{l(l+1)}{r^2} - 4\pi A^4(\varphi)\left(5\rho + 9p + \frac{(p+\rho)}{dp/d\rho}\right)\right) + \nu'^2\right)H = 0 \tag{26}$$

$$\delta\varphi'' + \left(\frac{2}{r} + \frac{1}{2}(\nu' - \lambda')\right)\delta\varphi' + e^{\lambda}\left(-\frac{l(l+1)}{r^2}\delta\varphi + 16A^3(\varphi)\pi\alpha(3p - \rho)\delta A + 4A^4(\varphi)\pi(3p - \rho)\delta\alpha\right) = 0, \tag{27}$$

where a prime ($'$) denotes the derivative with respect to $r$ and $\lambda$ refers to the metric function and not the tidal deformability. In the case of spontaneous scalarization, Equation (27) becomes

$$\delta\varphi'' + \left(\frac{2}{r} + \frac{1}{2}(\nu' - \lambda')\right)\delta\varphi' + e^{\lambda}\left(-\frac{l(l+1)}{r^2} + 4A^4(\varphi)\pi\beta(3p - \rho)(1 + 4\alpha\varphi)\right)\delta\varphi = 0. \tag{28}$$

External to the star, Equations (26) and (28), reduce to

$$H'' + \frac{2(r-\mu)}{r(r-2\mu)}H' - \left(e^{\lambda}\frac{l(l+1)}{r^2} + \left(\frac{2\mu}{r(r-2\mu)} + r\psi^2\right)^2\right)H = 0 \tag{29a}$$

[3] We thank Gastón Creci for his insights on this issue.

$$\delta\varphi'' + \frac{2(r-\mu)}{r(r-2\mu)}\delta\varphi' - \left(e^{\lambda}\frac{l(l+1)}{r^2}\right)\delta\varphi = 0. \tag{29b}$$

Equation (29(a)) depends on the $\psi$ and so is coupled to the scalar wave Equation 6b(b). As long as $\psi > 0$, there is no analytical solution to Equation (29(a)). Only approximate solutions exist at the surface of the star ($\tilde{p} = 0$) because $\psi \neq 0$. Since $\varphi$ asymptotically approaches a constant value $\varphi_\infty$, the derivative $\psi$ vanishes at large $r$. In this regime, Equation (29(a)) has an exact solution. The solutions to Equations 29(a), (b) are

$$H = c_1 Q_\ell^2\left(\frac{r}{\mu} - 1\right) + c_2 P_\ell^2\left(\frac{r}{\mu} - 1\right) \tag{30a}$$

$$\delta\varphi = d_1 Q_\ell\left(\frac{r}{\mu} - 1\right) + d_2 P_\ell\left(\frac{r}{\mu} - 1\right), \tag{30b}$$

where $P_\ell^m$ and $Q_\ell^m$ are the associated Legendre functions of the first and second kind.

In order to determine $c_1$, $c_2$, $d_1$, and $d_2$, we match the asymptotic behavior of the two solutions, i.e.,

$$H = \frac{8}{5}c_1\left(\frac{r}{\mu}\right)^{-3} + \mathcal{O}\left(\left(\frac{r}{\mu}\right)^{-4}\right) + 3c_2\left(\frac{r}{\mu}\right)^2 + \mathcal{O}\left(\frac{r}{\mu}\right); \tag{31a}$$

$$\delta\varphi = \frac{2}{15}d_1\left(\frac{r}{\mu}\right)^{-3} + \mathcal{O}\left(\left(\frac{r}{\mu}\right)^{-4}\right) + \frac{3}{2}d_2\left(\frac{r}{\mu}\right)^2 + \mathcal{O}\left(\frac{r}{\mu}\right) \tag{31b}$$

to the expansion of the $g_{tt}$ and the scalar field component of the metric (Equation (25)) respectively.

This gives us $c_1$, $c_2$, $d_1$, and $d_2$ in terms of the tensor tidal deformability $\lambda$ and the scalar tidal deformability $\lambda^{(\varphi)}$ respectively. For example, in the $\ell = 2$ case, we have

$$c_1 = \frac{15}{8}\frac{1}{\mu^3}\lambda\mathcal{E}, \quad c_2 = \frac{1}{3}\mu^2\mathcal{E}; \tag{32}$$

$$d_1 = \frac{45}{2}\frac{1}{\mu^3}\lambda^{(\varphi)}\mathcal{E}^{(\varphi)}, \quad d_2 = \frac{2}{3}\mu^2\mathcal{E}^{(\varphi)}. \tag{33}$$

By requiring continuity of the logarithmic derivatives

$$y = \frac{rH'}{H}, \text{ and } w = \frac{r\delta\varphi'}{\delta\varphi}, \tag{34}$$

and thus of $H$, $\delta\varphi$, and their derivatives at the surface of the star, it is possible to determine $\lambda$ and $\lambda^{(\varphi)}$ in terms of $\mu$, $r$, and either $y$ or $w$ respectively. This is done by substituting Equations (32) and (30(a)) or Equations (33) and (30(b)) into Equation (34) and solving for $\lambda$ or $\lambda^{(\varphi)}$.

The tidal Love numbers are connected to the tidal deformabilities by the following equations

$$k_\ell = \frac{(2\ell-1)!!}{2}\lambda R^{-(2\ell+1)}, \text{ and}$$
$$\kappa_\ell = \frac{(2\ell-1)!!}{2}\lambda^{(\varphi)} R^{-(2\ell+1)}. \tag{35}$$

Lastly, we can define the dimensionless tidal deformabilities:

$$\Lambda = \frac{\lambda}{\mu^{2\ell+1}}, \text{ and } \Lambda^{(\varphi)} = \frac{\lambda^{(\varphi)}}{\mu^{2\ell+1}}. \tag{36}$$

While this approach is sufficient to *define* $\lambda$, some difficulty arises when numerically calculating $\lambda$ and $k_\ell$ because Equation (30(a)) is only a solution to Equation (29(a)) in the large $r$ limit. It is not a solution near the surface of the star where numerical matching is typically done. In this region, an exact solution does not exist. While there is not an exact solution, an approximate series solution to Equation (29(a)) can be constructed order by order in powers of $r/\mu$. The leading order behavior of $H$ is

$$H \approx a_\ell\left(\frac{r}{\mu}\right)^\ell + a_{-(\ell+1)}\left(\frac{r}{\mu}\right)^{-(\ell+1)}. \tag{37}$$

However, the leading order solution alone is not accurate enough for our purposes. The solution to Equation (29(a)) is a linear superposition of the growing and diminishing solutions with two coefficients $a_\ell$ and $a_{-(\ell+1)}$, which are determined by the boundary conditions.

To create a more accurate solution, we construct two series solutions by adding higher order terms. There is one growing and one diminishing solution that correspond to the two terms in Equation (37). From there, higher order terms are added to construct a solution with the form

$$H \approx a_\ell\left(\left(\frac{r}{\mu}\right)^\ell + \alpha_1^+\left(\frac{r}{\mu}\right)^{\ell-1} + \alpha_2^+\left(\frac{r}{\mu}\right)^{\ell-2} + \ldots + \alpha_n^+\left(\frac{r}{\mu}\right)^{\ell-n}\right) + a_{-(\ell+1)}\left(\left(\frac{r}{\mu}\right)^{-(\ell+1)} + \alpha_1^-\left(\frac{r}{\mu}\right)^{-(\ell+2)} + \alpha_2^-\left(\frac{r}{\mu}\right)^{-(\ell+3)} + \ldots + \alpha_n^-\left(\frac{r}{\mu}\right)^{-(\ell+1+n)}\right). \tag{38}$$

For numerical purposes, the series is truncated at order $n = 13$. This ensures the series has converged within 0.5%.

Note that $H$ has only two degrees of freedom ($a_\ell$, $a_{-(\ell+1)}$). All other constants, $\alpha^+$ and $\alpha^-$, are functions of these two. The constants are determined by substituting one series solution, either growing or decaying, into Equation (29(a)) and solving for the coefficients order by order.

Lastly, by matching $a_\ell$ and $a_{-(\ell+1)}$ to Equation (25) and then substituting Equation (38) into Equation (34), the approximate tidal deformability $\lambda$ can be derived in the same method described above.

### 4.2. Magnetic Type Love Numbers

The odd parity or magnetic Love numbers $j_\ell$ and their associated tidal deformabilities $\sigma_\ell$ are functions of the odd parity metric perturbation $h_{\mu\nu}^-(h_0, h_1)$. The odd parity metric perturbations $h_0$ and $h_1$ (Equation (17)) are coupled only to the explicit fluid velocity perturbation $U(r)Y_{\ell m}$:

$$\delta u^{\mu} = [4\pi e^{-\nu/2} r^2 A^4(\varphi)(\rho + p)]^{-1}(0, 0, \partial_{\phi} Y_{\ell m}(\theta, \phi), \partial_{\theta} Y_{\ell m}(\theta, \phi))\sin^{-1}\theta\, U(r). \tag{39}$$

The odd parity metric perturbations can be constrained by three equations, which come from the $t\phi$, $r\phi$, and $\theta\phi$ components of the perturbation equations (see Equations (B27) to (B29)). There are multiple approaches to the magnetic Love number in the literature, but two are worthy of note (Pani et al. 2018). The earliest two publications on magnetic tidal deformabilities (Binnington & Poisson 2009; Damour & Nagar 2009) have approaches that are fundamentally different and whose results do not agree. The first approach developed by Binnington and Poisson (Binnington & Poisson 2009) assumes a *strictly static* fluid, i.e., $h_{0t} = h_{1t} = U = 0$. The second approach from Damour and Nagar (Damour & Nagar 2009) assumes an *irrotational* fluid. Instead of initially setting $h_{0t} = h_{1t} = U = 0$, this approach calculates the Regge–Wheeler equation and then takes the static limit ($\omega \to 0$). Note that these approaches seem equivalent at a surface level but do not lead to the same answer because the irrotational approach picks up a nonvanishing term from the fluid velocity perturbation. This section lays out both approaches and clarifies the subtle differences between them.

#### 4.2.1. Static Approach

In this section, we apply the static method derived in Binnington & Poisson (2009) to scalar-tensor theory; we assume that the perturbations are strictly static, i.e., $h_{0t} = h_{1t} = U = 0$, and consider the odd parity perturbation Equations (B27) to (B29). Under this assumption, Equation (B28) becomes

$$h_1 = 0, \tag{40}$$

and Equation (B29) becomes independent of $h_0$ and constrains only $h_1$.

The final remaining equation, Equation (B27), yields a second-order differential equation for $h_0$

$$e^{-\lambda}h''_0 - [4\pi r A^4(\varphi)(p + \rho) + e^{-\lambda} r \psi^2] h'_0 - \left[\frac{l(l+1)}{r^2} - \frac{4\mu}{r^3} + 8\pi A^4(\varphi)(p + \rho) - 2e^{-\lambda}\psi^2\right] h_0 = 0. \tag{41}$$

This equation is consistent with Equation (B7) in Sotani & Kokkotas (2005).

In the region exterior to the star, $\mu(r) = \mu$, and $\tilde{p} = \tilde{\rho} = 0$, and Equation (41) takes on a simpler form

$$e^{-\lambda}h''_0 - \left[\frac{l(l+1)}{r^2} - \frac{4\mu}{r^3} - 2e^{-\lambda}\psi^2\right] h_0 = 0. \tag{42}$$

This equation differs from the GR equation by the factor of $-2e^{-\lambda}\psi^2$ (Binnington & Poisson 2009). Equation (42) is now coupled to the scalar wave from Equation (6) and no longer has an exact solution. This result differs from the $f(R)$ results (Yazadjiev et al. 2018). The coupling functions $A(\varphi)$ differ between the two theories. Additionally, in $f(R)$ theories, $\varphi$ approaches zero as $r \to \infty$, whereas in the theories considered here $\varphi$ approaches a constant, nonzero value. Rather than matching solutions at the surface, Yazadjiev et al. (2018) matches the numerical solution to an analytical solution at some $r_{\rm match}$, beyond which the $\psi$ term can be neglected. $r_{\rm match}$ is defined by the Compton wavelength of the scalar field.

Since Equation (42) is true for all $r > r_s$, where $r_s$ is the surface of the star, including the large $r$ regime where $r \gg r_s$ and $\psi \to 0$, there is an exact solution in the large $r$ limit. This is sufficient to define the tidal Love numbers.

However, as was the case with the electric tidal deformability, there is no analytical solution at the surface of the star. Furthermore, the static approach is considered less physically relevant than the irrotational approach.

While an approximate solution to the static case could be constructed using the method laid out in Section 4.1 or a method similar to that used in the $f(R)$ case by Yazadjiev et al. (2018), this work focuses on the irrotational case because it is more realistic and has an analytical solution (Shapiro 1996; Landry & Poisson 2015; Pani et al. 2018).

#### 4.2.2. Irrotational Approach

In the *irrotational* approach, which was initially presented in Damour & Nagar (2009), it is assumed that the perturbations have a standard $e^{-i\omega t}$ time dependence, i.e., $h_i(r, t) = h_i(r)e^{-i\omega t}$.

Previous authors (Cunningham et al. 1978; Kojima 1992; Andrade & Price 1999; Damour & Nagar 2009) have noted that Equation (B29) can be solved for $h_0$ in terms of $h_1$ unless one assumes that $h_{0t} = 0$ (for that case, see Section 4.2.1).

Under this assumption, Equation (B29) can be rewritten as

$$h_{0t} = e^{(\nu-\lambda)/2}(\Psi r)'; \tag{43}$$

where $\Psi$ is defined such that

$$h_1 = e^{(\lambda-\nu)/2}\Psi r. \tag{44}$$

Assuming that $h_0(r, t) = h_0(r)e^{-i\omega t}$, Equation (43) can be used to define $h_0$:

$$h_0 = \frac{i}{\omega} e^{(\nu-\lambda)/2}(\Psi r)'. \tag{45}$$

It is evident from this equation that $h_0$ is not well defined in the $\omega \to 0$ limit (Pani et al. 2018). Substituting Equations (45) and (44) into Equation (B27) gives the following master equation:

$$\Psi'' + \frac{e^{\lambda}}{r^2}[2\mu + 4\pi r^3 A^4(\varphi)(p - \rho)]\Psi' + e^{\lambda}\left[e^{-\nu}\omega^2 + \frac{6\mu}{r^3} - \frac{l(l+1)}{r^2} + 4\pi A^4(\varphi)(p - \rho)\right] = 0. \tag{46}$$

This agrees with Equation (40) in Sotani & Kokkotas (2005).

Since it is assumed that the neutron star is static, we are interested in the $\omega \to 0$ limit. The master equation becomes continuous at the surface of the star, the logarithmic derivative $y^{odd} = r\psi'/\psi$ must also be continuous at the surface of the star.

$$\Psi'' + \frac{e^{\lambda}}{r^2}[2\mu + 4\pi r^3 A^4(\varphi)(p - \rho)]\Psi' + e^{\lambda}\left[\frac{6\mu}{r^3} - \frac{l(l+1)}{r^2} + 4\pi A^4(\varphi)(p - \rho)\right] = 0. \quad (47)$$

Outside of the star, this equation simplifies further:

$$\Psi'' + e^{\lambda}\frac{2\mu}{r^2}\Psi' + e^{\lambda}\left[\frac{6\mu}{r^3} - \frac{l(l+1)}{r^2}\right] = 0. \quad (48)$$

Interestingly, this equation, unlike the static master equation (Equation (41)), does not depend explicitly in $\varphi$ or $\psi$. Therefore, external to the star, the solution to Equation (47) is known and identical to the GR solution. All non-GR effects arise from matching the internal and external solutions at the surface of the star.

We briefly demonstrate the difference in the static and irrotational solutions in scalar-tenor theory. The method presented in Pani et al. (2018) is applied to the scalar-tensor problem.

Using the axial component of the stress-energy tensor conservation Equation (Equation (12)) and assuming $\omega \neq 0$, one finds that

$$U(r) = -4\pi A^3(\varphi)(\rho + p)e^{-\nu}h_0. \quad (49)$$

Substituting Equation (49) into Equation (B27) and then taking the $\omega \to 0$ limit, the following differential equation for $h_0$ is obtained

$$e^{-\lambda}h''_0 - [4\pi A^4(\varphi)(p + \rho)r + e^{-\lambda}r\psi^2]h'_0 - \left[\frac{l(l+1)}{r^2} - \frac{4\mu}{r^3} - 8\pi A^4(\varphi)(p + \rho) + 2e^{-\lambda}\varphi_r^2\right]h_0 = 0. \quad (50)$$

There is a sign change in the $8\pi A^4(\varphi)(p + \rho)$ term between Equation (41) and Equation (50). The difference occurs because $U(r) = 0$ for the static approach and $U(r) \neq 0$ in the irrotational case. So, while there is irrotational fluid motion in one case, the other has a completely static fluid. As a result of this difference, the irrotational tidal Love numbers are negative while the static Love numbers are positive.

Returning to the main goal of this work, calculating $j_\ell$, $\sigma_\ell$, and $\Sigma_\ell$, we use the similarity between the irrotational master equation and its GR counterpart to define the $\ell = 2$ solution. Equation (48) has an exact solution of the form

$$\Psi^{ext}(R) = b_p\Psi_p(R) + b_q\Psi_q(R) = b_pR^{\ell+1} - \frac{b_q}{4}R^3\left[R^{-4}F\left(\ell - 1, \ell + 2; 2\ell + 2; \frac{2}{R}\right)\right]; \quad (51)$$

where $R = r/\mu$, and $F$ is a hypergeometric function. For $\ell = 2$, $F$ is expressible in terms of simple functions.

$b_q$ and $b_p$ are determined by the boundary conditions at the surface of the star. Since both $\psi$ and $\psi'$ are required to be continuous at the surface of the star, the logarithmic derivative

$j_\ell$, $\sigma_\ell$, and $\Sigma_\ell$ are therefore defined to be

$$j_\ell \equiv -C^{2\ell+1}\frac{\psi'_p(R_s) - Cy^{odd}\psi_p(R_s)}{\psi'_q(R_s) - Cy^{odd}\psi_q(R_s)}, \quad (52)$$

$$\Sigma_\ell = \frac{\ell - 1}{4(l+2)}\frac{j_\ell}{(2\ell - 1)!!}C^{-(2\ell+1)}, \quad (53)$$

and

$$\sigma_\ell = \Sigma_\ell \mu^{2\ell+1} \quad (54)$$

where $R_s = r_s/\mu$, and $C$ is the compactness.

$j_\ell$ and $\sigma_\ell$ are, at a glance, identical to their GR counterparts, but the scalar-tensor and GR values differ because all non-GR effects are contained in the value of $y_s^{odd}$ calculated by integrating Equation (47) along with the modified TOV Equations (14a) to (14e) inside the star.

## 5. Results

### 5.1. Electric Love Numbers

This section presents the electric tidal Love numbers and the associated tidal deformabilites and compares them to the GR results. There are two degrees of freedom needed to define a specific case of spontaneous scalarization: $\beta$ and $\varphi_\infty$. $\beta$ is constrained by binary pulsar experiments to $\beta < -5$ at the $1\sigma$ level (Freire et al. 2012). In this work, we use several values of $\beta$ to demonstrate the results: $\beta = -4.5$, $-5$, $-5.5$, $-6$. Generally, the figures compare only $\beta = -4.5$ and $\beta = -6$. This gives two sets of results, one conservative and one optimistic. The value of the scalar field at infinity, $\varphi_\infty$, is tightly constrained by the Cassini experiment (Bertotti et al. 2003). That experiment directly constrains the Brans–Dicke parameter $\omega_{BD}$ to be $>4 \times 10^4$. The value of the scalar field at infinity is related to the Brans–Dicke parameter by the equation

$$\varphi_\infty = \frac{2}{|\beta|}\sqrt{\frac{\pi}{3 + 2\omega_{BD}}}. \quad (55)$$

This constrains $\varphi_\infty$ to $<2.7 \times 10^{-3}$ and $<2.0 \times 10^{-3}$ for $\beta = -4.5$, and $\beta = -6$ respectively. We use $\varphi_\infty = 10^{-3}$ for all results presented. Changing $\varphi_\infty$ to $2.0 \times 10^{-3}$ increases the deviation from GR. Conversely, changing $\varphi_\infty$ to $10^{-4}$ decreases the deviation from GR. These differences grow with increasing compactness but are less than 1% for the values considered.

Using Equation (30(a)) and Equation (35), it is possible to define the tidal Love number in the large $r$ limit. The scalar tidal Love number can be similarly calculated.

The $\ell = 2$ tidal Love numbers are defined as follows

$$k_2 = \frac{8(2C-1)^2C^5(2+2C(y-1)-y)}{5(2C(6+C^2(26-22y)-3y+4C^4(1+y)+3C(-8+5y)+C^3(6y-4))-3(1-2C)^2(2+2C(y-1)-y)\ln(1-2C))}; \quad (56a)$$

$$\kappa_2 = \frac{4C^5(2C-1)(2C^2w-6Cw+3w+6C-6)}{45(2C(C(12-9w)+3(-2+w)+C^2(-2+6w))+(-1+2C)(-6+6C+3w-6Cw+2C^2w)\ln[1-2C])}, \quad (56b)$$

where $y = rH'/H$, and $w = r\delta\varphi'/\delta\varphi$.

$y$ is traditionally evaluated at the star's surface for numerical applications. However, Equation (56(a)) is not valid when $r = r_s$. Close to the star $\psi \neq 0$, and the solution to Equation (29(a)) can only be approximated. After constructing a series solution that is accurate to better than 0.5% for even the largest values of $\psi$ considered (see Section 4.1), we compared values from the exact solution (Equation (56(a))) evaluated at the surface to the values of $k_2$ calculated from the approximate solution. The tidal deformabilities agreed to better than 3.7% for all equations of state and values of $\beta$ explored. The percent difference between the approximate and the exact values is strongly dependent on the compactness and increases with increasing compactness. For the vast majority of the parameter space explored, the difference between scalar-tensor theory and GR is larger than the difference between approximate and exact solutions. Exceptions occur for $\beta = -6$ where the scalar-tensor theory and GR curves intersect. This can be seen in Figure 1.

Figure 1 shows how the electric tidal Love numbers and tidal deformabilities differ in scalar-tensor theory and GR. Three different equations of state are considered: FPS, SLy, and MS1. These equations of state cover a wide range of stiffness and support a maximum mass of $>1.8M_\odot$. FPS and SLy are both within constraints from analyses of GW170817 (Abbott et al. 2017; Capano et al. 2020). However, as NICER results favor stiffer equations of state, we include MS1 (Bogdanov et al. 2019a, 2019b; Raaijmakers et al. 2020, 2021).

Figure 1 plots the physical or Jordan frame values, which are related to their Einstein frame counterparts by Equations (C18) and (C19). In Figures 1(a), (b), and (c), the observables $\tilde{\lambda}_2$, $\tilde{\Lambda}_2$, and $\tilde{k}_2$ are plotted against the neutron star's compactness ($\tilde{C}$), in this case defined as the Jordan frame TOV mass ($\tilde{M}$) over the Jordan frame radius $\tilde{r}_s$. In Figures 1(d), (e), and (f), the percent difference between scalar-tensor theory and GR is shown, also as a function of compactness. Note that Figures 1(e) and (f) are essentially identical. This is due to the definition of the dimensionless tidal deformability (Equation (36)). As $k_\ell$ and $\Lambda_\ell$ are related by a factor of $\frac{3}{2}\tilde{C}^5$ and $\tilde{C}$ is the $x$-axis variable, the factors of $\frac{3}{2}\tilde{C}^5$ cancel out. This can easily be shown by substituting the definition of the tidal Love number into the equation for the percent difference and forcing $C_{GR} = \tilde{C}$. This same phenomenon appears in Figure 3.

It is clear that the spontaneous scalarization effect can lead to significant deviations from the GR tidal deformabilities. It is also clear that the deviations are strongly dependent on the objects' compactness and the coupling constant. For the case where $\beta = -6$, the tidal Love number and dimensionless tidal deformability differ at most by ∼25%, and the tidal deformability differs by up to ∼200%. The peak occurs around $\tilde{C} \approx 0.25$ for the tidal deformability and ≈1.9 for the Love number, with the exact value varying by equation of state. In the more conservative case where $\beta = -4.5$, this reduces to ∼15% for the tidal Love number and ∼20% for the tidal deformability, and the peaks occur around $\tilde{C} \approx 0.3$ and ≈2.3 respectively.

The tidal deformability curve for scalar-tensor theories has a different shape than those in GR: a second peak appears. This peak is small for the weak coupling case, but for more negative coupling constants, the second peak is clear. This second peak is caused by the spontaneous scalarization effect, which causes large deviations from GR in conditions with strong gravitational fields (Damour & Esposito-Farèse 1993, 1996).

As the difference between scalar-tensor theory and GR is much greater than the difference between the two methods of calculating $k_\ell$, we consider Equation (56(a)) evaluated at the surface of the star to be sufficiently accurate for GW parameter estimation with current detectors.

We show the Jordan frame $\ell = 2$ scalar tidal Love numbers and tidal deformabilities in Figure 2. Scalar tidal Love numbers will effect scalar GW emission (Bernard 2020). We find that scalar tidal deformabilities are much smaller than the electric tidal deformabilities, around 2 orders of magnitudes smaller even for strongly scalarized cases. Additionally, the scalar tidal deformabilites and tidal Love numbers depend strongly on the coupling constant, with strong scalarization leading to negative scalar tidal Love numbers.

The $\ell = 3, 4$ tidal Love numbers are in Appendix D.

### 5.2. Magnetic Love Numbers

This section presents the magnetic tidal Love number and the associated tidal deformabilites in scalar-tensor theory and compares them to the GR results.

The exact equations for the magnetic tidal Love numbers $j_\ell$ and tidal deformabilities ($\sigma_\ell$ and $\Sigma_\ell$) can be determined by substituting Equation (51) into Equations (52), (53), and (54).

The explicit equation for the $\ell = 2$ or quadrupolar tidal Love number is

$$j_2 = \frac{96C^5(2C-1)(y-3)}{5(2C(12(y+1)C^4 + 2(y-3)C^3 + 2(y-3)C^2 + 3(y-3)C - 3y + 9) + 3(2C-1)(y-3)\ln(1-2C))}, \quad (57)$$

where $C = \mu/r_s$ is the Einstein frame compactness, and $y = y^{\text{odd}}(r_s) = r_s\Psi'/\Psi$ is the logarithmic derivative at the surface.

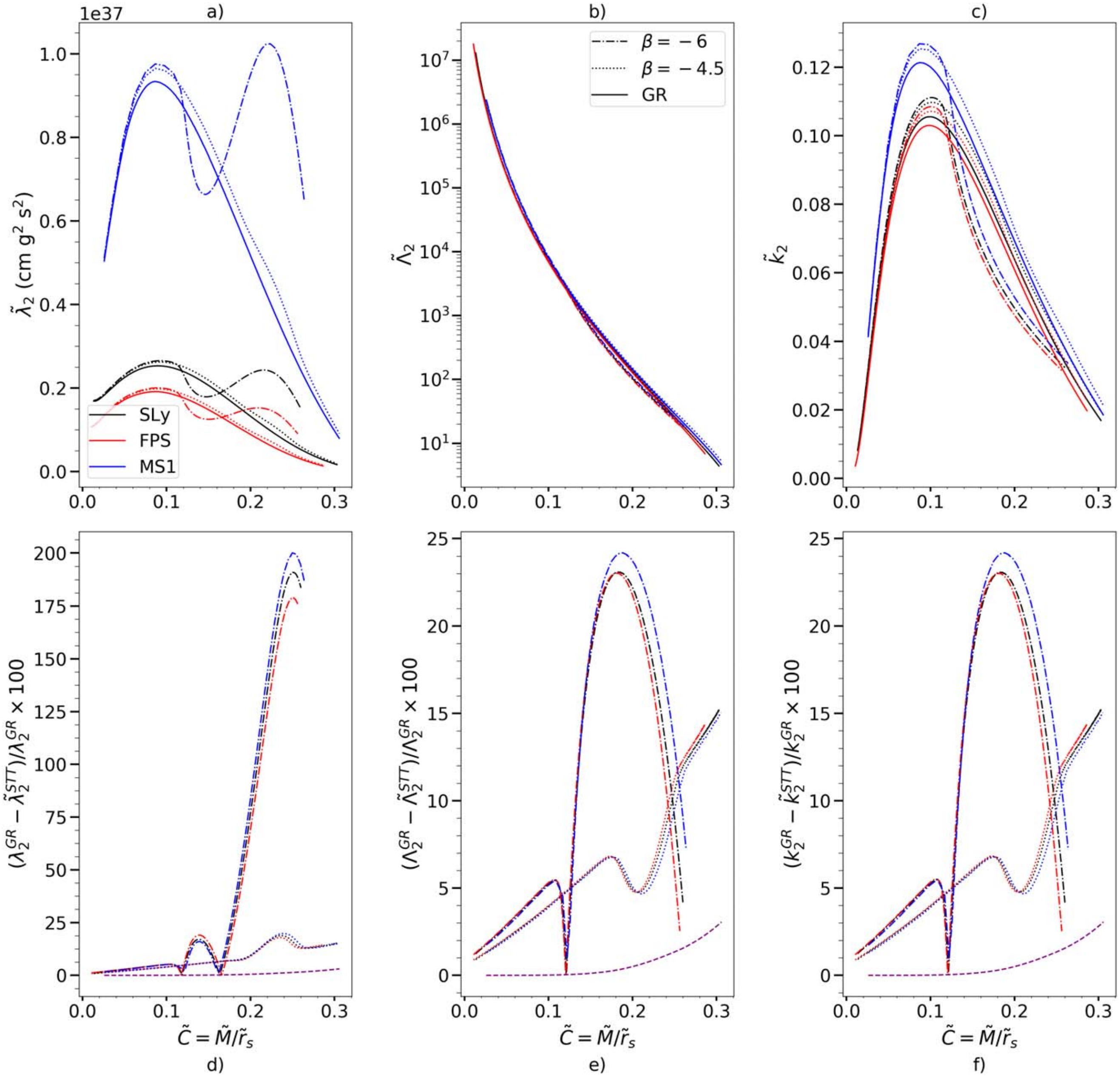

**Figure 1.** Panel (a) shows the $\ell = 2$ Jordan frame tidal deformability in cgs units, (b) shows the $\ell = 2$ Jordan frame dimensionless tidal deformability, (c) shows the $\ell = 2$ tidal Love number, (d) shows the percent difference between the cgs tidal deformability in scalar-tensor theory and GR, (e) shows the percent difference between the dimensionless tidal deformability in scalar-tensor theory and GR, and (f) shows the percent difference between the tidal Love numbers in scalar-tensor theory and GR. All are shown as a function of the Jordan frame compactness. The value of the scalar field at infinity $\varphi_\infty$ for all cases presented here is $10^{-3}$. Three realistic nuclear equations of state (SLy, FPS, and MS1) are shown in black, blue, and red respectively. The results for $\beta = -6$, and $\beta = -4.5$ are shown with dashed–dotted and dotted line styles. The purple lines in (d), (e), and (f) indicate the percent difference between the analytical and approximate approaches to calculating the tidal Love number and tidal deformability.

Figure 3 shows the Jordan frame $\ell = 2$ Love numbers, tidal deformabilities, and the difference between the GR and scalar-tensor tidal effects. The Jordan frame values are related to their Einstein frame counterparts by Equations (C10(a)–(b)). The conformal transformations are derived in Appendix C.

It is clear that tidal Love numbers and tidal deformabilities differ between GR and scalar-tensor theory. For the optimistic case where $\beta = -6$, the tidal Love number has a maximum deviation of $\sim$17%, and the tidal deformability has a maximum deviation of ~300%. This maximum deviation occurs at $\tilde{C} \approx 0.2$ for the Love number and ≈0.24 for the tidal deformability. In the more conservative case where $\beta = -4.5$, the peak occurs at $\tilde{C} \approx 0.23$ for all tidal properties, and the deviation changes to ~2.5% and ~15% for $j_2$ and $\sigma_2$ respectively.

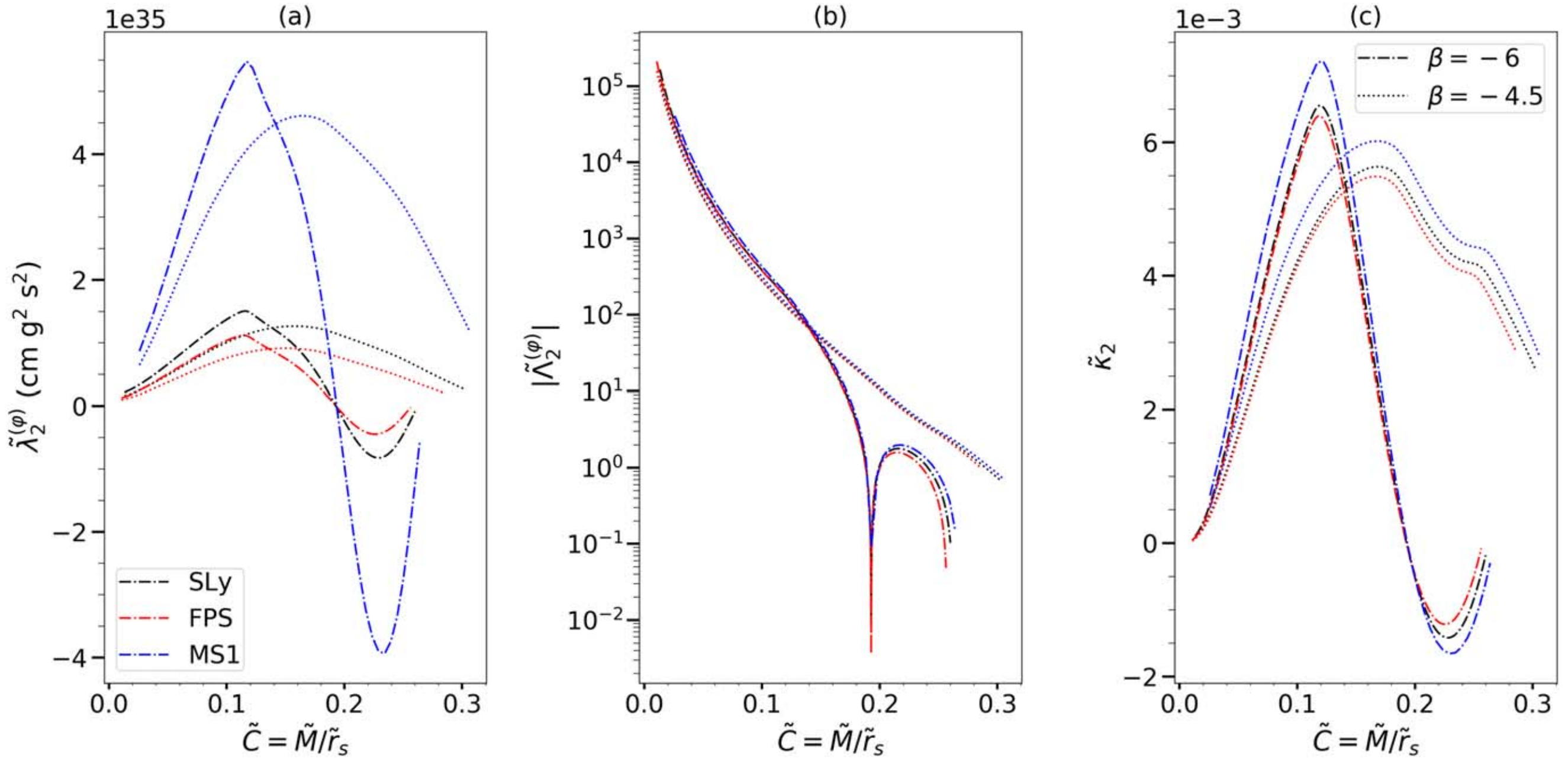

**Figure 2.** Panel (a) shows the $\ell = 2$ Jordan frame scalar tidal deformability in cgs units, (b) shows the $\ell = 2$ Jordan frame dimensionless scalar tidal deformability, (c) shows the scalar tidal Love number. All are shown as a function of the Jordan frame compactness. The value of the scalar field at infinity $\varphi_\infty$ for all cases presented here is $10^{-3}$. Three realistic nuclear equations of state (SLy , FPS, and MS1) are shown in black, red, and blue respectively. The results for $\beta = -6$, and $\beta = -4.5$ are shown with dashed–dotted and dotted line styles.

In GR, empirical relationships between the $\ell = 2$ dimensionless magnetic and electric tidal deformabilities have been found (Forteza et al. 2018). The dimensionless magnetic tidal deformability $\Sigma_2$ and the dimensionless electric tidal deformability $\Lambda_2$ have a quasi equation of state independent relationship:

$$\ln(-\Sigma_2) = \sum_{n=0}^{5} a_n (\ln\Lambda_2)^n . \tag{58}$$

We find that the scalar-tensor tidal deformabilities can be fit to a similar relationship, with the coefficients (written in Table 1) depending on the value of $\beta$. This relationship is shown for several values of $\beta$ in Figure 4. Regardless of equation of state, $R^2 > 0.99$ for all cases.

There does not appear to be a similar relationship between the scalar deformability and the electric tidal deformability. The scalar tidal deformability depends strongly on $\beta$ and $\Lambda^{(\varphi)}/\Lambda$ can take on different shapes that are equation of state dependent.

## 6. Discussion

This work presents the electric, magnetic, and scalar tidal Love numbers and tidal deformabilities. We find that the electric and magnetic tidal effects may differ significantly from their general relativistic counterparts (~200 and ~300 for electric and magnetic respectively). These large deviations occur at larger compactnesses ($\gtrsim$2) and are caused by the spontaneous scalarization effect. The exact deviation and the compactness where this maximum deviation occurs are equation of state dependent.

This paper approaches tidal effects through the lens of GW parameter estimation. The mass–radius–tidal deformability relationships explored in this paper can be applied directly to GW parameter estimation of GWs from binary neutron star and neutron star black hole systems. The $\ell = 2$ dimensionless electric tidal deformability is the leading order tidal effect for GWs. Given that this number can vary by ~25 between scalar-tensor theory and GR, it may be necessary to take modified tidal effects into account when doing tests of GR using GWs from systems with neutron stars.

We present an analytical expression for the magnetic tidal Love numbers in scalar-tensor theory for the first time. The results establish that the magnetic Love numbers are only implicitly dependent on the scalar field and have an analytical solution. This is in agreement with Sotani & Kokkotas (2005), which shows that the time-dependent perturbation equation is only implicitly dependent on the scalar field. However, this was discussed only in the context of perturbations and not of tidal Love numbers.

The magnetic Love numbers in this paper can be compared to their $f(R)$ counterparts because $f(R)$ theory and scalar-tensor theory are mathematically similar. Differences arise when calculating the tidal Love numbers in part due to the behavior of the scalar field at infinity. In $f(R)$ theory, the scalar field and its derivative go to zero at infinity, and the tidal deformability can be evaluated at some distance away from the neutron star where both the scalar field and its derivative are sufficiently small. This is different from the scalar-tensor theories considered in this paper where the scalar field asymptotically approaches a constant. Additionally, the coupling function differs between theories with $A(\varphi) \propto e^{\alpha\varphi}$ in $f(R)$. Despite this, the perturbation equations inside the star should agree when the correct substitutions for $A(\varphi)$ and $\alpha(\varphi)$ have been made because they are mathematically similar. However, our perturbation equation differs from Equation (23) in Yazadjiev et al. (2018) by a negative sign. Comparing the tidal Love numbers themselves, shown in Figure 3, with the results from Yazadjiev et al. (2018), it is clear that the qualitative features are consistent, with the deviation from GR increasing with compactness. However, the difference between GR and scalar-tensor theory are smaller than those between GR and $f(R)$, at least for physically allowed values of $\beta$ and $\varphi_\infty$.

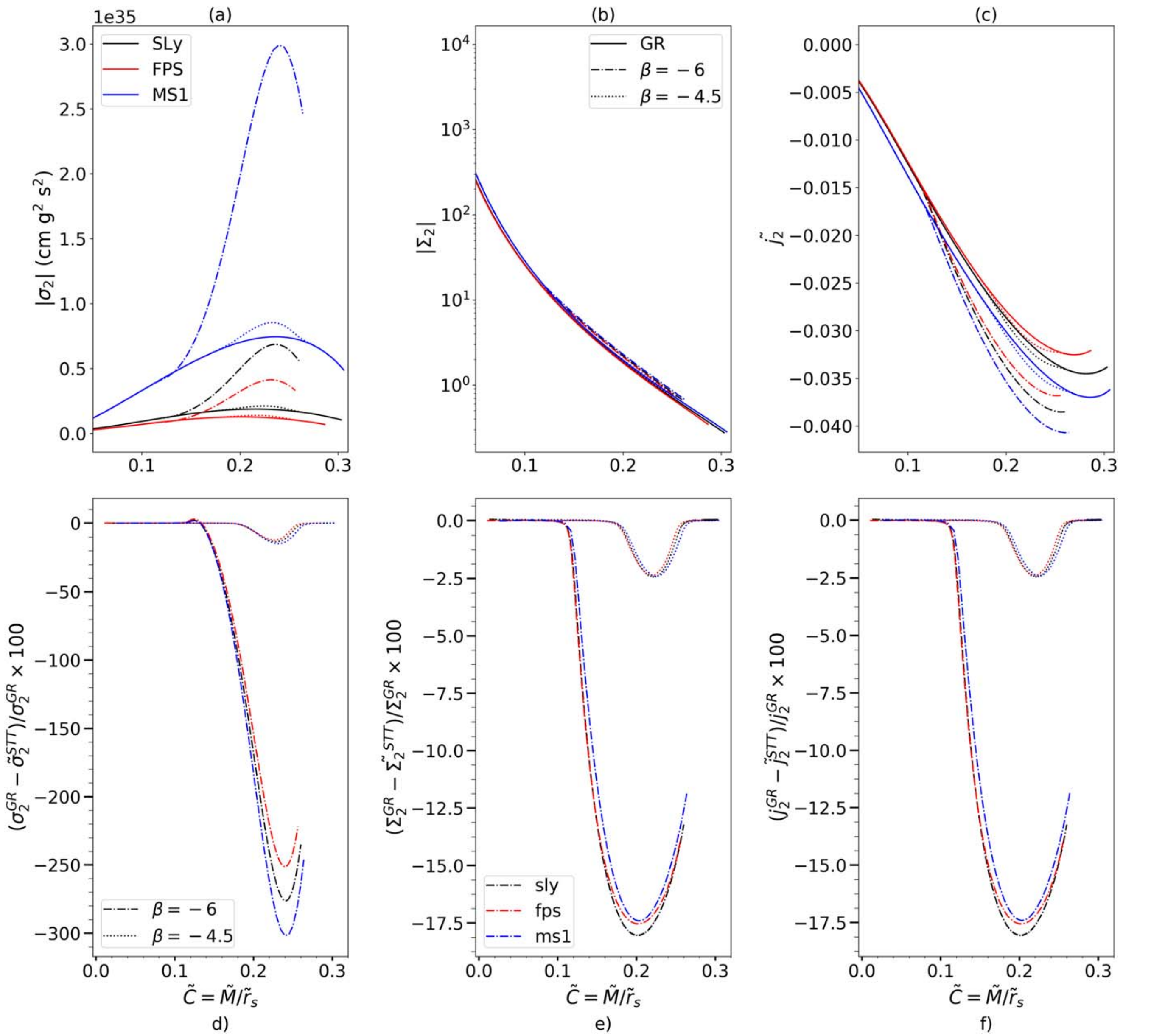

**Figure 3.** Panel (a) shows the $\ell = 2$ Jordan frame tidal deformability in cgs units, (b) shows the $\ell = 2$ Jordan frame dimensionless tidal deformability, (c) shows the tidal Love number, (d) shows the percent difference between the cgs tidal deformability in scalar-tensor theory and GR, (e) shows the percent difference between the dimensionless tidal deformability in scalar-tensor theory and GR, and (f) shows the percent difference between the tidal Love numbers in scalar-tensor theory and GR. All are shown as a function of the Jordan frame compactness. The value of the scalar field at infinity $\varphi_\infty$ for all cases presented here is $10^{-3}$. Three realistic nuclear equations of state (SLy, FPS, and MS1) are shown in black, red, and blue respectively. The results for $\beta = -6$, and $\beta = -4.5$ are shown with dashed–dotted and dotted line styles.

This paper also includes the even parity tidal Love numbers and tidal deformabilities. The $\ell = 2$ electric tidal Love numbers in scalar-tensor theory were initially presented in Pani & Berti (2014) in the context of the so-called "I-Love-Q" relations. The methods in this paper differ significantly from those in Pani & Berti (2014).

To begin, Pani & Berti (2014) use the most general stationary axisymmetric metric that includes first-order rotation terms rather than the stationary Schwarzshield metric used in this paper. In addition to this, there is a fundamental difference between the definitions of the tidal Love numbers and tidal deformabilities between this work and theirs. This is based on the way that the even parity perturbation equations are treated. There are both metric and scalar perturbations in the even parity case, and the relationship between them is not trivial. In the Einstein frame, the metric tensor and the scalar field are not coupled. As a change in the metric should not affect the scalar field and vice versa, it is important to construct two independent first-order perturbation equations, one for the metric perturbation and one for the scalar. This differs from the approach in Pani & Berti (2014), where the two even parity equations are coupled. It is unsurprising, then, that Equations (26) and (27) are different from the equations presented in Pani & Berti (2014). The resulting tidal Love numbers must also differ. The definition for the scalar tidal Love number in this paper also differs from that in Pani & Berti (2014). Pani & Berti (2014) does not include a source term in their definition of the scalar tidal Love number, and we do. This is because they are considering the perturbation in the scalar field produced by a change in the metric rather than by a change in the scalar field. This paper also includes the $\ell = 3$, 4 even parity Love numbers and tidal deformabilities in Appendix D, which have not been presented before.

**Table 1**
Fit Coefficients

| Theory | $a_0$ | $a_1$ | $a_2$ | $a_3$ | $a_4$ | $a_5$ |
|---|---|---|---|---|---|---|
| GR | −1.99 | $4.51 \times 10^{-1}$ | $1.60 \times 10^{-2}$ | $6.51 \times 10^{-4}$ | $-1.07 \times 10^{-4}$ | $3.74 \times 10^{-6}$ |
| $\beta = -4.5$ | 1.13 | −3.29 | 1.68 | $-3.48 \times 10^{-1}$ | $3.48 \times 10^{-2}$ | $-1.34 \times 10^{-3}$ |
| $\beta = -5$ | −2.53 | $7.44 \times 10^{-1}$ | $-7.85 \times 10^{-2}$ | $3.39 \times 10^{-2}$ | $-6.07 \times 10^{-3}$ | $3.61 \times 10^{-4}$ |
| $\beta = -5.5$ | −4.24 | 2.53 | $-7.60 \times 10^{-1}$ | $1.56 \times 10^{-1}$ | $-1.60 \times 10^{-2}$ | $6.41 \times 10^{-4}$ |
| $\beta = -6$ | −3.55 | 1.76 | $-3.80 \times 10^{-1}$ | $6.73 \times 10^{-2}$ | $-6.08 \times 10^{-3}$ | $2.10 \times 10^{-4}$ |

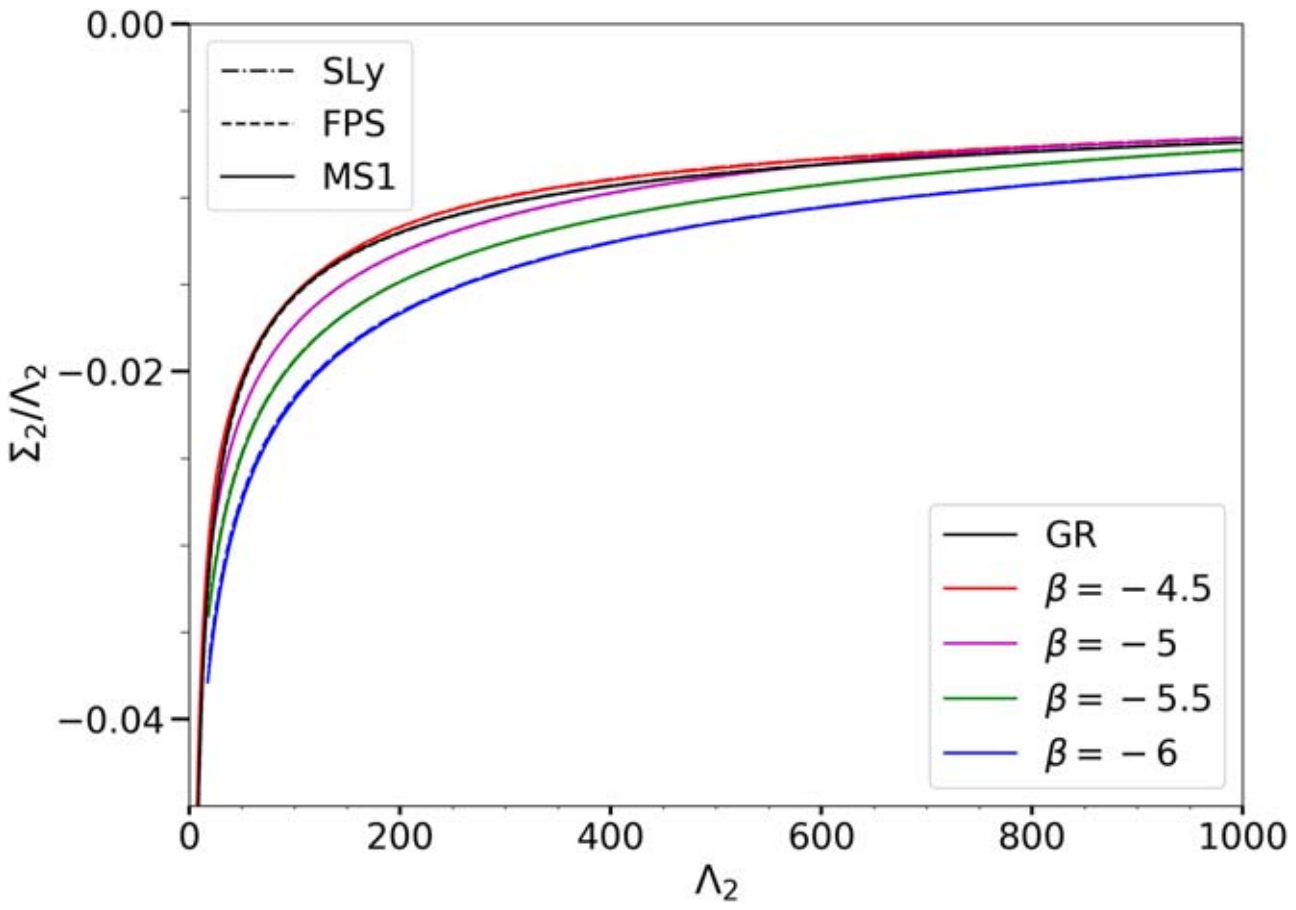

**Figure 4.** Quasiuniversal relations between the Jordan frame dimensionless magnetic quadrupolar tidal deformability ($\Sigma_2$) and the Jordan frame dimensionless electric quadrupolar tidal deformability $\Lambda_2$. Only the irrotational magnetic tidal deformability is shown in this figure. Three equations of state (SLy, FPS, and MS1) are shown in different line styles, but they are indistinguishable. The color of the lines corresponds to different values of $\beta$.

The results demonstrate that tidal Love numbers and tidal deformabilities can differ significantly between scalar-tensor theory and GR. This is consistent with other results in the literature, which show that tidal Love numbers in $f(R)$ theory and scalar-Gauss–Bonnet gravity (Yazadjiev et al. 2018; Saffer & Yagi 2021) also differ significantly from their general relativistic counterparts. As GWs emitted by neutron stars depend on the tidal deformability, it is essential to take the changes in the mass, radius, and tidal deformability into account when studying GWs from neutron stars in theories beyond GR. The allowed deviations from GR in the GWs are smaller than or similar to the uncertainty in the tidal deformability measurements from binary neutron star and neutron star black hole mergers. By taking the modified tidal deformability into account, the small deviations from GR in the waveform can be more accurately determined.

## Acknowledgments

We thank Badri Krishnan, Xisco Jiménez Forteza, Sayak Datta, Sumit Kumar, Pierre Mourier, and Gaston Creci for their valuable discussions. Our computations used the ATLAS computing cluster at AEI Hannover funded by the Max Planck Society and the State of Niedersachsen, Germany.

## Appendix A
## Numerical Methods

This section lays out the numerical techniques used to calculate the mass, radius, and tidal deformability relations shown in Section 5, Figures 1–2.

First, the structure equations presented in Section 2 are solved numerically using Scipy's solve_ivp with the "DOP853" option, which is an eighth-order Runge–Kutta method. To validate the results, the "DOP853" results are compared to those from the older odeint solver and solve_ivp's "RK45" option, which is a Runge–Kutta solver that uses a fifth-order accurate formula but calculates the accuracy using the fourth-order method (Virtanen et al. 2020). Next, the perturbation equations are added to the TOV solver, and the tidal deformabilities and Love numbers are calculated from Equations (52), (56(a)), and (56(b)).

As discussed in Section 2, in scalar-tensor theories, the structure equations can be expressed in either the Jordan frame or the Einstein frame. The code takes advantage of the relative simplicity of the Einstein frame structure equations to numerically construct the neutron star model. The Jordan frame quantities are calculated at the end of the code, using the Einstein frame values and the conformal transformations in Appendix C.

### *A.1. Background Configuration*

For numerical integration, the equations need to be posed as first-order ordinary differential equations of the form

$$y_i'(x) = f(x, y_i, y_i'); \tag{A1}$$

where $x = r$ is the independent variable, and $y = M$, $\nu$, $\varphi$, $\psi$, and $p$ are the dependent variables.

It is important to consider that the values of $r$, $M$, $\nu$, $\psi$, and $p$ may vary greatly in magnitude, which can lead to numerical errors and instabilities. Codes often use scale factors to mitigate the numerical errors. In Manolidis (2014), it is claimed that by choosing $\rho = \rho_0 \hat{\rho}$, $p = \rho_0 \hat{p}$, $M = r_0 \hat{M}$, and $r = r_0 \hat{r}$ the form of the modified TOV equations remains unchanged so long as $\rho_0 r_0^2 = 1$. However, this is not the case.

This code uses a new set of scale factors. Specifically, we scale only $\tilde{p}$ and $\tilde{p}$ and not $M$ or $r$. With $\rho = \tilde{\rho}_0\hat{\rho}$, and $p = \tilde{p}_0\hat{p}$, we are able to lay out the structure equations used in the code:

$$\frac{d\mu}{dr} = 4\pi G_* r^2 A^4(\varphi)\tilde{\rho}_0\hat{\rho} + \frac{1}{2}r(r-2\mu)\psi^2; \tag{A2a}$$

$$\frac{d\nu}{dr} = 8\pi G_* \frac{r^2 A^4(\varphi)\tilde{p}_0\hat{p}}{r-2\mu} + r\psi^2 + \frac{2\mu}{r(r-2\mu)}; \tag{A2b}$$

$$\frac{d\varphi}{dr} = \psi; \tag{A2c}$$

$$\frac{d\psi}{dr} = 4\pi G_* \frac{rA^4(\varphi)}{r-2\mu}[\alpha(\varphi)(\tilde{\rho}_0\hat{\rho} - 3\tilde{p}_0\hat{p}) + r\psi(\tilde{\rho}_0\hat{\rho} - \tilde{p}_0\hat{p})] - \frac{2(r-\mu)}{r(r-2\mu)}\psi; \tag{A2d}$$

$$\frac{d\hat{p}}{dr} = -\frac{1}{\tilde{p}_0}(\tilde{\rho}_0\hat{\rho} + \tilde{p}_0\hat{p})\left[4\pi G_* \frac{r^2 A^4(\varphi)\tilde{p}_0\hat{p}}{r-2\mu} + \frac{1}{2}r\psi^2 + \frac{\mu}{r(r-2\mu)} + \alpha(\varphi)\psi\right]. \tag{A2e}$$

In order to solve these equations, the numerical solver requires initial conditions. In this case, the initial conditions are defined near the center of the star ($r \approx 0$). Due to numerical instabilities at $r = 0$, the code starts at some small, but finite radius (e.g., $r_0 = 10^{-5}$ m). We used a convergence test to ensure that $r_0$ was sufficiently small and would not effect the final results.

We know that

$$\mu(r=0) = 0; \tag{A3a}$$

$$\nu(r=0) = 0; \tag{A3b}$$

$$\psi(r=0) = 0; \tag{A3c}$$

and the initial pressure $\tilde{p}$ varies. However, the scalar field is defined at infinity $\varphi(\infty) = \varphi_0$, and not at $r = 0$. The shooting method is employed to convert the boundary value problem into an initial value problem.

The process begins with an initial guess for $\varphi(r=0) = \varphi_c$. The system of equations is then integrated outward to the star's surface, which is defined to be where the pressure vanishes ($\tilde{p} = 0$). The code then calculates the value of the scalar field at infinity $\varphi_\infty$ using the relationship between $\varphi_s$, the value of $\varphi$ at the surface, and $\varphi_\infty$. The connection between $\varphi_s$ and $\varphi_\infty$ can be found by solving the scalar wave equation outside of the star and matching the interior and exterior solutions:

$$\varphi_\infty = \varphi_s + \frac{2\psi_s}{\sqrt{\nu_s'^2 + 4\psi_s^2}} \operatorname{arctanh}\left(\frac{\sqrt{\nu_s'^2 + 4\psi_s^2}}{\nu'_s + 2/r_s}\right), \tag{A4}$$

where subscript $s$ indicates values evaluated at the surface, and the prime ($'$) denotes derivative with respect to $r$.

The code then compares the calculated value of the scalar field at infinity $\varphi_\infty$ to the actual value of the scalar field at infinity $\varphi_0$. The parameter $\Delta\varphi = \varphi_\infty - \varphi_c$ is calculated, and if $\Delta\varphi$ is greater than some tolerance (here, $\Delta\varphi \leqslant 10^{-5}$), then the value of $\varphi_c$ is updated, and the process is repeated. The process is repeated until $\varphi_\infty$ agrees with $\varphi_0$ within some tolerance. For a more in depth discussion on the shooting method, see, for instance, Press et al. (2007).

In order to solve the TOV equations, it is necessary to provide an equation of state $\tilde{p}(\tilde{\rho})$, which relates the Jordan frame pressure and density. In this work, we consider a variety of equations of state. All equations of state are defined in the physical frame. In order to include realistic equations of state, our code takes in equation of state data from external data files. The code obtains the density at any point by taking the given pressure and the data from the file and interpolating.

The SLy (Douchin & Haensel 2001), FPS (Friedman & Pandharipande 1981), and MS1 (Müller & Serot 1996) equation of state are considered because they are commonly used in literature and useful for comparison with previous results (Lattimer & Prakash 2001; Read et al. 2009).

### A.2. Tidal Deformability

The definitions of the tidal deformabilites were derived in Section 5, and now, we focus on calculating them. First, we must integrate the perturbation equations for $H$, $\Psi$, and $\delta\varphi$ along with the scalar-tensor TOV equations. The initial value problem solver requires that we recast the second-order differential Equations (26), (28), and (47) into first-order differential equations. There are two ways to do this. One, any second-order differential equation can be recast as a system of two first-order differential equations. Two, a single first-order differential equation for the logarithmic derivative (e.g., $y = rH'/H$) can be obtained from the original equation. As the definitions of the tidal deformabilities and tidal Love numbers from Equations (52), (56)(a)–(b)) depend on the logarithmic derivative, we recast Equations (26), (28), and (47) into first-order differential equations for the logarithmic derivative. These equations now have form:

$$\frac{dy(r)}{dr} = -\frac{1}{r}(y^2(r) + y(r)F(r) + r^2 Q(r)). \tag{A5}$$

For the magnetic perturbations,

$$F(r) = \left(1 - \frac{2\mu}{r}\right)^{-1}\left(\frac{2\mu}{r} + 4\pi A^4(\varphi)r^2(\tilde{p} - \tilde{\rho})\right) - 1; \tag{A6a}$$

$$r^2Q(r) = \left(1 - \frac{2\mu}{r}\right)^{-1}\left(\frac{\ell(\ell+1)\mu}{r} + 4\pi A^4(\varphi)r^2(\tilde{p} - \tilde{\rho}) - 6\right). \tag{A6b}$$

For the even parity tensor perturbations,

$$F(r) = \left(1 - \frac{2\mu}{r}\right)^{-1}(1 + 4\pi A^4(\varphi)r^2(\tilde{p} - \tilde{\rho})); \tag{A7a}$$

$$r^2Q(r) = \left(1 - \frac{2\mu}{r}\right)^{-1}\left(-\ell(\ell+1) + \frac{4\pi A^4(\varphi)r^2(\tilde{p} + \tilde{\rho})}{dp/d\rho} + 4\pi A^4(\varphi)r^2(9\tilde{p} + 5\tilde{\rho})\right) - \left(\left(1 - \frac{2\mu}{r}\right)^{-1}(2\mu + 8\pi A^4(\varphi)r^3\tilde{p}) + r^3\psi^2\right)^2. \tag{A7b}$$

For the scalar perturbations,

$$F(r) = \left(1 - \frac{2\mu}{r}\right)^{-1}(1 + 4\pi A^4(\varphi)r^2(\tilde{p} - \tilde{\rho})); \tag{A8a}$$

$$r^2Q(r) = \left(1 - \frac{2\mu}{r}\right)^{-1}(4\pi A^4(\varphi)r^2(1 + 4\varphi\alpha(\varphi))\beta(3\tilde{p} - \tilde{\rho}) - \ell(\ell+1))). \tag{A8b}$$

The initial conditions are

$$y_{\text{even}}(r=0) = 2; \tag{A9a}$$

$$y_{\text{scalar}}(r=0) = 2; \tag{A9b}$$

$$y_{\text{odd}}(r=0) = 3. \tag{A9c}$$

The values of $y_{\text{odd}}$, $y_{\text{even}}$, and $y_{\text{scalar}}$ at the surface are then determined, and the Love numbers can be calculated. Lastly, the Jordan frame values are calculated using then conformal transformations derived in Appendix C.

## Appendix B
## Perturbation Equations

### B.1. Perturbed Energy-momentum Tensor

In this section of the appendix, the exact forms of the fluid stress-energy tensor perturbations are given. Subscripts are used to denote derivatives.

The pressure and density perturbations are defined in the physical frame to be $\delta\tilde{p}(r)Y_{\ell m}$ and $\delta\tilde{\rho}(r)Y_{\ell m}$. The fluid velocity and its perturbations are also written in the Jordan frame. In the case of static tides, the fluid velocity perturbation is generally a function only of the metric perturbations and does not have explicit velocity perturbations. Furthermore, as the tides are static, the total perturbed four-velocity has the following form:

$$\hat{u}^\mu = u^\mu + \delta u^\mu = (\hat{u}^0, 0, 0, 0) \tag{B1}$$

where the tilde has been dropped for readability. The time component of $\hat{u}^\mu$ differs from $u^\mu$ because the perturbed metric differs from the unperturbed metric.

The even parity velocity perturbations are

$$\delta\tilde{u}^t = \frac{1}{2A(\varphi)}e^{-\nu/2}H_0Y_{\ell m}; \tag{B2a}$$

$$\delta\tilde{u}^i = 0, \quad i = 1, 2, 3. \tag{B2b}$$

While time-independent perturbations do not depend explicitly on fluid velocity perturbations, the time-dependent equations do. The two methods presented in Section 4.2 differ in the way that the fluid velocity term $U(r)$ is treated. In both cases, the explicit dependence vanishes, but their results differ because of how they treat this term. The time-dependent odd parity velocity perturbations are

$$\delta\tilde{u}^\phi = \frac{e^{\nu/2}U(r)e^{-i\omega t}}{4\pi A^4(\varphi)(\tilde{p} + \tilde{\rho})}\csc\theta\partial_\theta Y_{\ell m}; \tag{B3a}$$

$$\delta\tilde{u}^\nu = 0, \quad \nu = 0, 1, 2. \tag{B3b}$$

The components of $\hat{u}_\mu$ are calculated by lowering the contravariant four-velocity $\hat{u}^\mu$ with the total metric $g_{\mu\nu} = (g^0_{\mu\nu} + h_{\mu\nu})$.

Additionally, the perturbed matter stress-energy tensor depends on the Eulerian fluid perturbations: $\delta\tilde{\rho}(r)Y_{\ell m}$ and $\delta\tilde{p}(r)Y_{\ell m}$ respectively. We assume a barotropic equation of state, and so

$$\delta\tilde{\rho} = \frac{\partial\tilde{\rho}}{\partial\tilde{p}}\delta\tilde{p}. \tag{B4}$$

Using these definitions and assuming that by symmetry $\partial_\phi Y_{\ell m} = 0$, the nonzero components of the perturbed matter stress-energy tensor are as follows:

$$\delta T^t_t = -(4A^3(\varphi)\tilde{\rho}\delta A + A^4(\varphi)\delta\tilde{\rho})Y_{\ell m}; \tag{B5}$$

$$\delta T^t_\phi = -\left(A^4(\varphi)\tilde{\rho}h_0 + \frac{A(\varphi)}{4\pi}e^\nu U\right)\sin\theta\partial_\theta Y_\ell; \tag{B6}$$

$$\delta T^r_r = (4A^3(\varphi)\tilde{p}\delta A + A^4(\varphi)\delta\tilde{p})Y_{\ell m}; \tag{B7}$$

$$\delta T^r_\phi = A^4(\varphi)\tilde{p}\ h_1\sin\theta\partial_\theta Y_{\ell m}; \tag{B8}$$

$$\delta T^\theta_\theta = (4A^3(\varphi)\tilde{p}\delta A + A^4(\varphi)\delta\tilde{p})Y_{\ell m}; \tag{B9}$$

$$\delta T^\phi_t = -\left(A^4(\varphi)\tilde{\rho}h_0 + \frac{A(\varphi)}{4\pi}e^\nu U\right)\sin\theta\partial_\theta Y_\ell; \tag{B10}$$

$$\delta T_r^{\phi} = A^4(\varphi)\tilde{p}\ h_1 \sin\theta\partial_\theta Y_{\ell m}; \quad \text{(B11)}$$

$$\delta T_\phi^{\phi} = (4A^3(\varphi)\tilde{p}\delta A + A^4(\varphi)\delta\tilde{p})Y_{\ell m}. \quad \text{(B12)}$$

The nonzero components of the perturbed energy-momentum tensor for the scalar field $T_{\mu\nu}^{(\varphi)}$ have the following form:

$$\delta T_{00}^{(\varphi)} = -2e^{\nu-\lambda}[H\psi^2 - \psi\delta\varphi']Y_{\ell m}; \quad \text{(B13)}$$

$$\delta T_{03}^{(\varphi)} = -e^{\lambda}\psi^2 h_0 \sin\theta\partial_\theta Y_{\ell m}; \quad \text{(B14)}$$

$$\delta T_{11}^{(\varphi)} = 2\psi\delta\varphi' Y_{\ell m}; \quad \text{(B15)}$$

$$\delta T_{12}^{(\varphi)} = 2\psi\delta\varphi\partial_\theta Y_{\ell m}; \quad \text{(B16)}$$

$$\delta T_{13}^{(\varphi)} = -e^{-\lambda}\psi^2 h_1 \sin\theta\partial_\theta Y_{\ell m}; \quad \text{(B17)}$$

$$\delta T_{22}^{(\varphi)} = r^2 e^{-\lambda}[(H-K)\psi^2 - 2\psi\delta\varphi']Y_{\ell m}; \quad \text{(B18)}$$

$$\delta T_{33}^{(\varphi)} = r^2 e^{-\lambda}[(H-K)\psi^2 - 2\psi\delta\varphi']\sin^2\theta\ Y_{\ell m}. \quad \text{(B19)}$$

### B.2. Equations for Even Parity

The following equations are derived from the even parity metric perturbation equations. The first six come from perturbing the Einstein equation (Equation (6)):

1. Equation (B20) is $\delta G_2^2 - \delta G_3^3 = 8\pi G_*(\delta T_2^2 - \delta T_3^3) + (\delta T_2^{(\varphi)2} - \delta T_3^{(\varphi)3})$;
2. Equation (B21) is $\delta G_1^2 = 8\pi G_*\delta T_1^2 + \delta T_1^{(\varphi)2}$;
3. Equation (B22) is $\partial_r(\delta G_1^2 = 8\pi G_*\delta T_1^2 + \delta T_1^{(\varphi)2})$;
4. Equation (B23) is $\delta G_1^1 = 8\pi G_*\delta T_1^1 + \delta T_1^{(\varphi)1}$;
5. Equation (B24) is $\delta G_2^2 + \delta G_3^3 = 8\pi G_*(\delta T_2^2 + \delta T_3^3) + (\delta T_2^{(\varphi)2} + \delta T_3^{(\varphi)3})$;
6. Equation (B25) is $\delta G_0^0 - \delta G_1^1 = 8\pi G_*(\delta T_0^0 - \delta T_1^1) + (\delta T_0^{(\varphi)0} - \delta T_1^{(\varphi)1})$.

$$H_0 = H_2. \quad \text{(B20)}$$

$$K' = H_{0,r} + \nu' H_0. \quad \text{(B21)}$$

$$K'' = H''_0 + \nu'' H_0 + \nu' H_{0,r}. \quad \text{(B22)}$$

The equation for the scalar perturbation $\delta\varphi$ is derived by perturbing scalar wave equation Equation (B26)).

$$\delta\varphi'' = \left(\frac{-4 + r\lambda' - r\nu'}{2r}\right)\delta\varphi' + e^{\lambda}\frac{l(l+1)}{r^2}\delta\varphi + 16\pi A^3(\varphi)e^{\lambda}\alpha(\tilde{\rho} - 3\tilde{p})\delta A + 4\pi A^4(\varphi)e^{\lambda}(\tilde{\rho} - 3\tilde{p})\delta\alpha. \quad \text{(B26)}$$

### B.3. Equations for Odd Parity

The static and irrotational methods used in this paper differ in their treatment of time derivatives. Even though the tidal Love numbers themselves are time-independent, we present the time-dependent equations in this section.

Combining the $\delta G_{\mu\nu}$ with the matter stress-energy tensor and scalar stress-energy tensor terms results in the following three equations:

1. Equation (B27) is $\delta G_{t\phi} = 8\pi\delta T_{t\phi} + T_{t\phi}^{(\varphi)}$;
2. Equation (B28) is $\delta G_{r\phi} = 8\pi\delta T_{r\phi} + T_{r\phi}^{(\varphi)}$;
3. Equation (B29) is $\delta G_{\theta\phi} = 8\pi\delta T_{\theta\phi} + T_{\theta\phi}^{(\varphi)}$.

$$e^{-\lambda}(h_{0rr} - h_{1rt}) - \frac{2e^{-\lambda}}{r}h_{1t} + [4\pi rA^4(\varphi)(p+\rho) + e^{-\lambda}r\psi^2](h_{1t} - h_{0r}) - \frac{1}{r^3}[l(l+1)r - 4m + 8\pi A^4(\varphi)(p+\rho)r^3 - 2r^3e^{-\lambda}\psi^2]h_0 - 4A(\varphi)e^{\nu}U = 0. \quad \text{(B27)}$$

$$e^{-\nu}(h_{0rt} - h_{1tt}) - 2\frac{e^{-\nu}}{r}h_{0t} - \left[\frac{l(l+1)-2}{r^2}\right]h_1 = 0. \quad \text{(B28)}$$

$$e^{-\nu}h_{0t} - \frac{1}{r^2}[2m - 4\pi r^3 A^4(\varphi)(\rho - p)]h_1 - e^{-\lambda}h_{1r} = 0. \quad \text{(B29)}$$

## Appendix C
## Conformal Transformations

The tidal Love numbers in this paper were derived in the Einstein frame; however, as experiments measure Jordan frame quantities, it is necessary to obtain the Jordan frame quantities using a conformal transformation. We assume here that the

$$(l(l+1)+2)K = (l(l+1) - 2e^{-\lambda}(1 + r\nu' - r^2\psi^2))H_0 - 2e^{-\lambda}rH_{0r} + e^{-\lambda}r(2 + r\nu')K' - 16\pi A^4(\varphi)r^2\delta\tilde{p}. \quad \text{(B23)}$$

$$16\pi A^4(\varphi)r^2\delta\tilde{p} = e^{-\lambda}\left(\frac{-4 + r\lambda' - 3r\nu'}{2r}\right)H_{0r} + e^{-\lambda}\left(\frac{4 - r\lambda' + r\nu'}{2r}\right)K' - e^{-\lambda}H_{0rr} + e^{-\lambda}K'' - e^{-\lambda}\left(\frac{4r\psi^2 - (\lambda' - \nu')(2 + r\nu') + 2r\nu''}{2r}\right). \quad \text{(B24)}$$

$$e^{-\lambda}K'' - e^{-\lambda}\left(\frac{-4 + r(\lambda' + \nu')}{2r}\right)K' + \left(\frac{e^{-\lambda}r(\lambda' + \nu' - 2r\psi^2) - l(l+1)}{r^2}\right)H_0 + 8\pi A^4(\varphi)\left(1 + \frac{d\rho}{dp}\right)\delta p = 0. \quad \text{(B25)}$$

Jordan frame metric $\tilde{g}_{\mu\nu}$ is related to the Einstein frame metric $g_{\mu\nu}$ by a conformal factor $A(\varphi)$:

$$\tilde{g}_{\mu\nu} = A^2(\varphi) g_{\mu\nu}, \tag{C1}$$

where $A(\varphi) = e^{\frac{1}{2}\beta\varphi^2}$. By construction, the Einstein frame metric is asymptotically flat. This implies that

$$\tilde{g}_{\mu\nu} \to A^2(\varphi)\eta_{\mu\nu} \tag{C2}$$

where $\eta_{\mu\nu} = \mathrm{diag}(-1, 1, 1, 1)$ is the Minkowski metric. As the Jordan frame metric is also asymptotically flat or Minkowskian, the $\tilde{r}$ and $\tilde{t}$ components must be related to their Einstein frame counterparts in the following way: $\tilde{r} = A(\varphi)r$, and $\tilde{t} = A(\varphi)t$. Furthermore, the effective gravitational constant $\tilde{G}$ is no longer a constant in the Jordan frame and is not necessarily equal to the bare gravitational constant $G$, which appears in the Einstein frame equations. The relationship between the two is known (Palenzuela et al. 2014):

$$\tilde{G} = e^{\beta\varphi_\infty^2}\left[G + \frac{\beta\varphi_\infty^2}{4\pi}\right]. \tag{C3}$$

We need the conformal transformations for the perturbations between the two frames to transform the tidal Love numbers and tidal deformabilities from the Einstein frame to the Jordan frame. These are presented in Section 3.

### C.1. Odd Parity

The odd parity perturbation in the Einstein frame $h_0$ is related to the odd parity perturbation in the Jordan frame by

$$\tilde{h}_0 = A^2(\varphi) h_0. \tag{C4}$$

To see how $\Psi$ transforms, it is easiest to start with the definition of $\Psi$ given in Damour & Nagar (2009):

$$\Psi = r^3 \partial_r\left(\frac{h_0}{r^2}\right) = r{h'}_0 - 2h_0. \tag{C5}$$

From this definition of $\Psi$, it is straightforward to show that it transforms as

$$\tilde{\Psi}(\tilde{r}) = A^2(\varphi)\Psi(r). \tag{C6}$$

To properly define the magnetic tidal deformability in the Jordan frame, $\tilde{\Psi}$ must have the same leading order behavior as $\Psi$, i.e.,

$$\tilde{\Psi}^{ext}(\tilde{R}) = \tilde{b}_p \tilde{R}^{\ell+1} + \tilde{b}_q \tilde{R}^{-\ell}; \tag{C7}$$

where $\tilde{R} = \tilde{r}/\tilde{\mu} = A^2(\varphi) r/\mu$. Equations (C7) and (C6) can be used to relate $\tilde{b}_{q,p}$ to their Einstein frame counterparts:

$$\tilde{b}_p = A^{-2\ell}(\varphi) b_p; \quad \tilde{b}_q = A^{2\ell+2}(\varphi) b_q. \tag{C8}$$

The Jordan frame tidal Love number $\tilde{j}_\ell$ is defined to be

$$\tilde{j}_\ell = \tilde{C}^{2\ell+1}\frac{\tilde{b}_q}{\tilde{b}_q}. \tag{C9}$$

From Equations (C8) and (C9), it follows that

$$\tilde{j}_\ell = j_\ell; \tag{C10a}$$

$$\tilde{\sigma}_\ell = (A^2(\varphi_\infty))^{2\ell+1}\sigma_\ell. \tag{C10b}$$

### C.2. Even Parity

To transform the scalar Love number between frames, it is only necessary to know the relationship between the scalar field in the Einstein ($\varphi$) and Jordan frames ($\phi$)

$$\phi = e^{-\beta\varphi^2}. \tag{C11}$$

By perturbing this equation, the relationship between the Jordan frame tidal deformability $\lambda^{(\phi)}$ and its Einstein frame counterpart $\lambda^{(\varphi)}$ can be derived:

$$\lambda^{(\phi)} = (A(\varphi_\infty))^{2\ell+1}\lambda^{(\varphi)}. \tag{C12}$$

The tidal Love numbers are related by

$$\kappa_\phi = \kappa_\varphi. \tag{C13}$$

In the case of the even parity tensor tidal Love number, the transformation between frames is more complex due to mixing of the scalar and tensor perturbations. The relationship between the even parity metric perturbations in the two frames is constrained by the choice of gauge. Taking the equation relating the time–time component of the metric perturbation in the Jordan frame $\tilde{H}$ to the Einstein frame metric perturbation $H$ and the Einstein frame scalar perturbation $\delta\varphi$ from Section 3, we have

$$\tilde{H} = A^2(\varphi) H - 2A(\varphi)\delta A. \tag{C14}$$

In the spontaneous scalarization case, this becomes

$$\tilde{H} = A^2(\varphi)(H - 2\beta\varphi\delta\varphi). \tag{C15}$$

Combining this with the leading order behavior of the perturbations, which are known to be

$$\begin{aligned} H &= -\mathcal{E}_{ij} r^2 + \mathcal{O}(r) + \frac{3Q_{ij}}{r^3} + \mathcal{O}(r^{-4}) \\ &= -\mathcal{E}_{ij} r^2 + \mathcal{O}(r) - \frac{3\lambda\mathcal{E}_{ij}}{r^3} + \mathcal{O}(r^{-4}), \end{aligned} \tag{C16}$$

it is possible to define the Jordan frame tidal deformabilty $\tilde{\lambda}_J$:

$$\tilde{\lambda}_J = \frac{A^{2\ell+1}(\varphi_\infty)(\lambda_E \mathcal{E}_{ij}^E - 2\beta\varphi_\infty \lambda^{(\varphi)}\mathcal{E}_{ij}^\varphi)}{\mathcal{E}_{ij}^E - 2\beta\varphi_\infty \mathcal{E}_{ij}^\varphi} \tag{C17}$$

where $E$ denotes Einstein frame tensor quantities, and $\varphi$ denotes Einstein frame scalar quantities. From this equation, it is clear that the Jordan frame tidal deformability is related linearly to the even parity scalar and tensor tidal deformabilities. The exact relationship is

$$\tilde{\lambda}_J = A^{2\ell+1}(\varphi_\infty)(\lambda_E + \lambda^{(\varphi)}). \tag{C18}$$

Finally, we determine the tidal Love numbers to have the following relationship

$$\tilde{k}_\ell = k_\ell + \kappa_\ell. \tag{C19}$$

## Appendix D
## Higher Order Love Numbers

Using Equations (30)(a)–(b)), and the methods presented in Section 4.1, we determine the equations for the $\ell = 3$, 4 tidal Love numbers and tidal deformabilities at large $r$.

The $\ell = 3$, 4 even parity tensor tidal Love numbers are defined as

$$k_3 = 8(1-2C)^2C^7(-3-3C(-2+y)+2C^2(-1+y)+y) \times [7(2C(15(-3+y)+4C^5(1+y)-45C(-5+2y)-20C^3(-9+7y)+2C^4(-2+9y) +5C^2(-72+37y))+15(1-2C)^2(-3-3C(-2+y)+2C^2(-1+y)+y)\ln(1-2C))]^{-1}, \tag{D1}$$

$$k_4 = 32(1-2C)^2C^9(-7(-4+y)+28C(-3+y)-34C^2(-2+y)+12C^3(-1+y)) \times [147(2C(C^2(5360-1910y)+C^4(1284-996y)-105(-4+y) + 8C^6(1+y)+105C(-24+7y)+40C^3(-116+55y)+C^5(-8+68y)) +15(1-2C)^2(-7(-4+y)+28C(-3+y)-34C^2(-2+y)+12C^3(-1+y))\ln(1-2C))]^{-1}, \tag{D2}$$

and the scalar tidal Love numbers are

$$\kappa_3 = 12C^7(-1+2C)(-5(-3+w)+15C(-2+w)-12C^2(-1+w)+2C^3w) \times[175(-2C(C^2(96-71w)-15(-3+w)+15C(-9+4w)+C^3(-6+22w)) +3(-1+2C)(-5(-3+w)+15C(-2+w)-12C^2(-1+w)+2C^3w)\ln(1-2C))]^{-1}, \tag{D3}$$

$$\kappa_4 = (64C^9(-1+2C)(35(-4+w)-140C(-3+w)+180C^2(-2+w)-80C^3(-1+w)+8C^4w)) \times[3675(-2C(105(-4+w)-190C^3(-4+3w)-105C(-16+5w)+4C^4(-6+25w)+10C^2(-206+89w)) +3(-1+2C)(35(-4+w)-140C(-3+w)+180C^2(-2+w)-80C^3(-1+w)+8C^4w)\ln(1-2C))]^{-1}. \tag{D4}$$

## ORCID iDs

Stephanie M. Brown https://orcid.org/0000-0003-2111-048X